\documentclass[
 reprint,
 APS
 amsmath,amssymb,
]{revtex4-2}

\usepackage{graphicx}% Include figure files
\usepackage{dcolumn}% Align table columns on decimal point
\usepackage{bm}% bold math
\usepackage[utf8]{inputenc}
\usepackage[T1]{fontenc}
\usepackage{mathptmx}
\usepackage{etoolbox}
\usepackage[toc,page]{appendix}
\usepackage{amsmath}
\usepackage[inkscapelatex=false]{svg}

\makeatletter
\def\@email#1#2{%
 \endgroup
 \patchcmd{\titleblock@produce}
  {\frontmatter@RRAPformat}
  {\frontmatter@RRAPformat{\produce@RRAP{*#1\href{mailto:#2}{#2}}}\frontmatter@RRAPformat}
  {}{}
}%
\makeatother

\newcommand{\etal}{\textit{et al.} }
\begin{document}

\preprint{AIP/123-QED}

%\title[]{Unclocklike biological oscillators with frequency memory}

\title[]{Unclocklike oscillators with frequency memory for the entrainment of biological clocks}

% Force line breaks with \\
\author{Christian Mauffette Denis}
\affiliation{%
D\'epartement de Physique, Universit\'e de Montr\'eal, Montr\'eal, QC, Canada}
\author{Paul François}%
 \email{paul.francois@umontreal.ca}
\affiliation{%
D\'epartement de Biochimie et Medecine Mol\'eculaire, Universit\'e de Montr\'eal, Montr\'eal, QC, Canada\\
 MILA Qu\'ebec, Montr\'eal, QC, Canada 
}
 %\email{Second.Author@institution.edu.}

\date{\today}% It is always \today, today,

\begin{abstract}
Entrainment experiments on the vertebrate segmentation clock have revealed that embryonic oscillators actively change their internal frequency to adapt to the driving signal. This is neither consistent with a one-dimensional clock model nor with a limit-cycle model, but rather suggests a new "unclocklike" behavior. In this work, we propose simple, biologically realistic descriptions of such internal frequency adaptation, where a phase oscillator activates a memory variable controlling the oscillator's frequency. We study two opposite limits for the control of the memory variable, one with a smooth phase-averaging memory field, and the other with a pulsatile, phase-dependent activation. Both models recapitulate intriguing properties of the entrained segmentation clock, such as very broad Arnold tongues and an entrainment phase plateauing with detuning. We compute analytically multiple properties of such systems, such as entrainment phases and cycle shapes. We further describe new phenomena, including hysteresis in entrainment, bistability in the frequency of the entrained oscillator, and probabilistic entrainment. Our work shows that oscillators with frequency memory can exhibit new classes of unclocklike properties, that can be tested through experimental entrainment.
\end{abstract}

\maketitle

%\begin{quotation}
%The ``lead paragraph'' is encapsulated with the \LaTeX\ 
%\verb+quotation+ environment and is formatted as a single paragraph before the first section heading. 
%(The \verb+quotation+ environment reverts to its usual meaning after the first sectioning command.) 
%Note that numbered references are allowed in the lead paragraph.
%
%The lead paragraph will only be found in an article being prepared for the journal \textit{Chaos}.
%\end{quotation}

\section{Introduction}

Non-linear oscillators are often coupled to external signals, giving rise to complex dynamical patterns and computations \cite{glass2001}. Multiple examples are found in cellular biology, e.g. the internal circadian clocks entrained by the day/night cycles \cite{partch2014} or the cell cycle itself \cite{novak2008}. Coupled cellular oscillations also play a major role during embryonic development, e.g. in the so-called "segmentation clock" \cite{Pourquie2022}. Segmentation in vertebrates is the periodic process leading to the formation of somites, which are vertebrae precursors. In 1976, Cooke and Zeeman proposed that the formation of somites was under the control of a global embryonic oscillator \cite{Cooke1976}, and in 1997, Palmeirm \etal confirmed how oscillations and spatial waves of genes implicated in the Notch signaling pathway were coupled to somite formation \cite{Palmeirim1997}. Since then, multiple experimental works have explored the properties of the segmentation clock \cite{oates2012, Pourquie2022}. The embryonic segmentation clock emerges from the coupling of numerous individual cellular oscillators, each of them displaying complex dynamics with a period slowing down \cite{Morelli2009, ares2012, Lauschke2013, Webb2016, Uriu2021}, see also \cite{francois2024} for an extensive review of existing mathematical models. 

\begin{figure*}
\includegraphics[width=18cm]{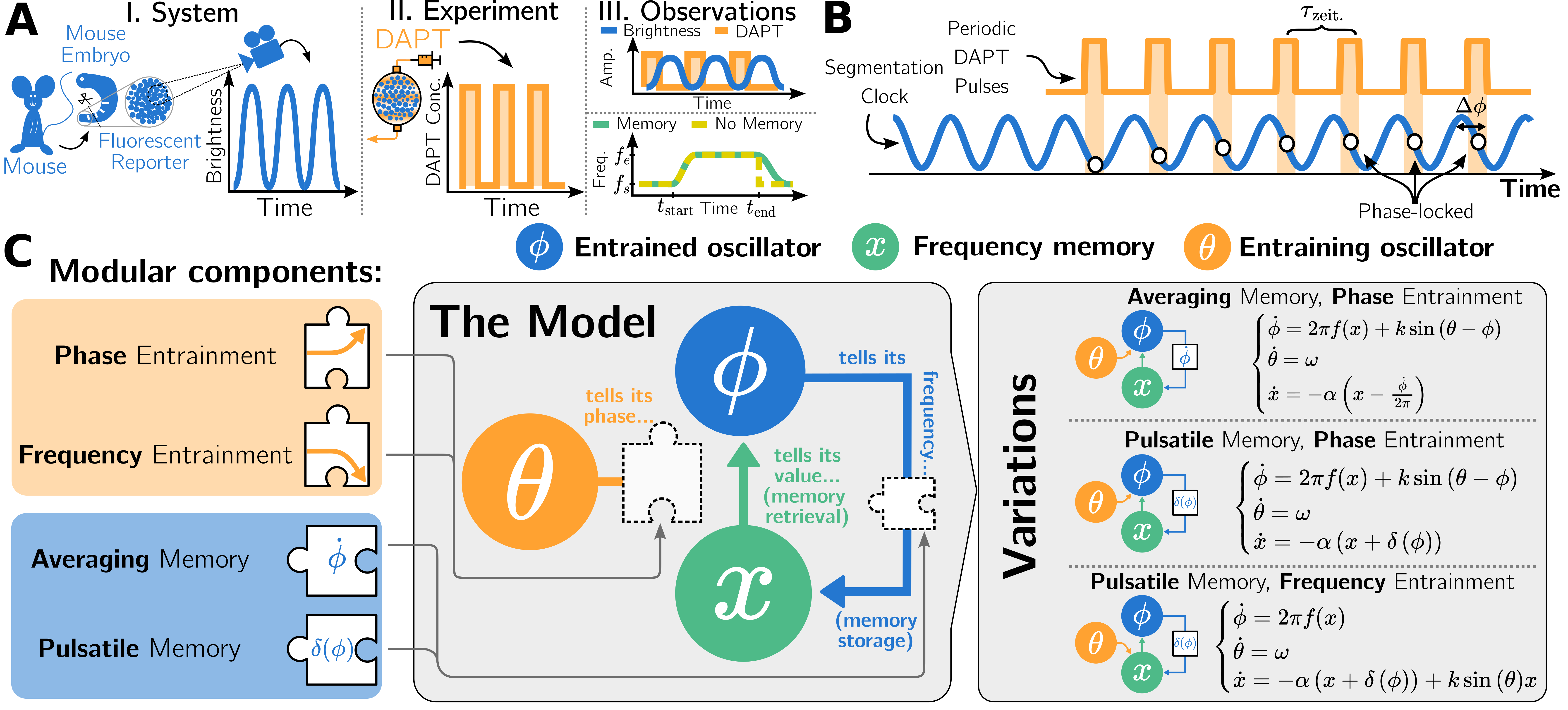}% Here is how to import EPS art
\caption{A) Sanchez et al \cite{Sanchez2022} studied oscillations in the Notch signaling pathway in the mouse embryo tail bud. Such oscillations are recorded through a dynamic Notch reporter called LuVeLu, accounting for Lfng oscillations (Panel I). Using a microfluidic chip, they periodically pulsated DAPT (a Notch inhibitor) into the system (Panel II). This external perturbation managed to entrain the Notch oscillations, over a broad range of periods (from 120 to 180 minutes, while the mouse's intrinsic segmentation clock's period is 140 minutes) (Panel III, top). Experiments, where the system was "released" from the entraining signal, also revealed that the embryonic oscillator adapts its internal frequency since it takes multiple cycles to go back to the original intrinsic frequency of 140 minutes. B) Periodic perturbation of the segmentation clock via DAPT (period $\tau_\text{zeit.}$) in presomitic tissue results in entrainment of the clock with measurable entrainment phase, $\Delta \phi$. C) The model we study has three interconnected dynamical variables. They are the entrained oscillator (blue), the entraining oscillator (orange), and the memory variable $x$ (green). The types of interconnections are modular and give rise to the variations of the model.} \label{fig:fig1}
\end{figure*}

Segmentation oscillators are now routinely cultured in the lab, where oscillations can be maintained for multiple cycles and visualized using standard fluorescent reporters in tissue culture \cite{Lauschke2013,hubaud2017}, stem cells systems \cite{Matsuda2020,miao2023} or even single cells \cite{Rohde2021}. Recently, Sanchez et al. \cite{Sanchez2022} further studied the entrainment properties of the segmentation oscillator submitted to periodic pulses of DAPT (a Notch inhibitor), Fig. \ref{fig:fig1} A-B. While the segmentation clock (measured via global fluorescence in the culture) can indeed be entrained, it was shown that the oscillator response unexpectedly depends on the period of the entrainment system: phase response curves computed at different periods are "shifted" with respect to one another, suggestive of internal period changes. Indeed, once DAPT pulses are stopped, it takes several cycles for the segmentation oscillator to relax to its initial intrinsic frequency (Fig. \ref{fig:fig1} A). 

 Such dependency on the entrainment period is not consistent with the classical theory for oscillator entrainment \cite{Winfree, Kuramoto, cross2005, Granada2013}. To see why, it is useful to introduce Winfree's distinctions between "clock" and "unclock" properties for biological oscillators \cite{Winfree1975}. "Clock" refers to systems where all accessible states lie on a circle, so that the dynamics are effectively 1D periodic, and the rates of transitions between states set timescales. Examples include early models of cell-cycle \cite{Winfree1975}, or quadratic-and-integrate fire neural models for neuroscience\cite{Izhikevich}, and such 1D cycle structure is proposed to provide cycle robustness \cite{li2004}. Perturbations of clocks should then move the system around its 1D periodic coordinate, i.e. modulate rates of transitions, or define a simple phase response curve \cite{Kuramoto, Winfree, cross2005}. Thus, we do not expect any dependency on the period of the entrainment system in the absence of any extra (internal) states. Going further, Winfree noticed that limit cycle oscillators are "unclocklike" because one can define and observe biologically functional states outside the cycle (e.g. the phaseless center of the cycle, associated with Type 0 resetting \cite{Winfree}). However, a dependency on the period of the entrainment signal is not consistent with classical entrainment theory \cite{Kuramoto, Winfree, Izhikevich} of limit cycles either (see Appendix \ref{app:classical_entrainment} for an explanation). 
 
 Thus, to explain a dependency of entrainment properties on the period of the external signal, we need new hypotheses. A possibility, observed in so-called "overdrive suppression" on cardiac oscillators, is that an external perturbation modifies chemical parameters driving the period \cite{Zheng2015}. Then, changes in oscillator properties with the entrainment period might appear incidental with no obvious biological meaning. However, in highly regulated processes such as embryonic development, changes in internal properties in response to signals might rather be associated with internal buffering or adaptive mechanisms (and associated internal states), e.g., regulating the developmental "tempo" \cite{manser2023, EBISUYA2024}. Winfree already mentioned the possibility of "clockshop" behaviors, where multiple oscillators would interact, giving rise to new "unclocklike" properties. Such interactions have already been studied for the coupling of two internal independent oscillators (e.g. cell-cycle and circadian clock \cite{feillet2014,droin2019} or master/slave circadian neurons \cite{jeong2022}), but importantly did not explicitly include the possibility of internal period changes. A "clockshop" scenario is very plausible for the segmentation clock oscillator(s) since multiple internal oscillators are implicated in this system \cite{Sonnen2018, Goldbeter2008}, with an already established functional role of oscillators' interaction for the slowing down of the dynamics and differentiation \cite{Lauschke2013, Sonnen2018}.

%Those experimental results suggest that the response function $H$, and by extension the limit cycle of the entrained oscillator, depends on the frequency of entrainment $\omega$.

%Such dependency has been observed in other contexts, e.g. on cardiac oscillators with the overdrive suppression in Calcium signaling \cite{Zheng2015}, which in turn leads to phase response curves depending on the stimulus in a non-trivial way. Overdrive suppression might correspond to a passive, undesired, dependency of the frequency to an external perturbation, but a 

 Here we revisit those ideas, to propose and study a class of minimal models with two oscillating variables accounting for unusual entrainment properties similar to the segmentation clock. In brief, we start with the simplest possible continuous clock (a phase oscillator $\phi$) and add one extra degree of freedom with feedback (a memory $x$). This renders the system bidimensional, and "unclocklike" in a different way from both the standard limit cycle models \cite{Winfree} or from the coupling of two phase oscillators \cite{feillet2014,droin2019}. We study two types of internal coupling, and derive conditions for stability, which are biologically accessible (see Fig. \ref{fig:fig1} C for a graphical summary of the models studied). Adding such memory variable can lead to \textit{frequency adaptation}, where an entrained oscillator can change its internal frequency to match a periodic stimulus. We show that such memory variable comes with specific entrainment properties. We then establish that this class of models recapitulates very well the unusual properties observed in the entrained segmentation clock such as a plateau in the entrainment phase as a function of detuning, and are well captured by simple, intuitive analytical expressions capturing the non-linearity of memory computation. Furthermore, we predict multiple new properties of such systems, such as the bistability of the internal frequency, that can not be accounted for by classical entrainment theory (i.e. without internal memory). While our work is theoretical, we discuss simple experimental protocols to probe such properties.

\section{Results}

\subsection{Modelling phase oscillators with frequency memory}

\begin{figure}
\includegraphics[width=8.5cm]{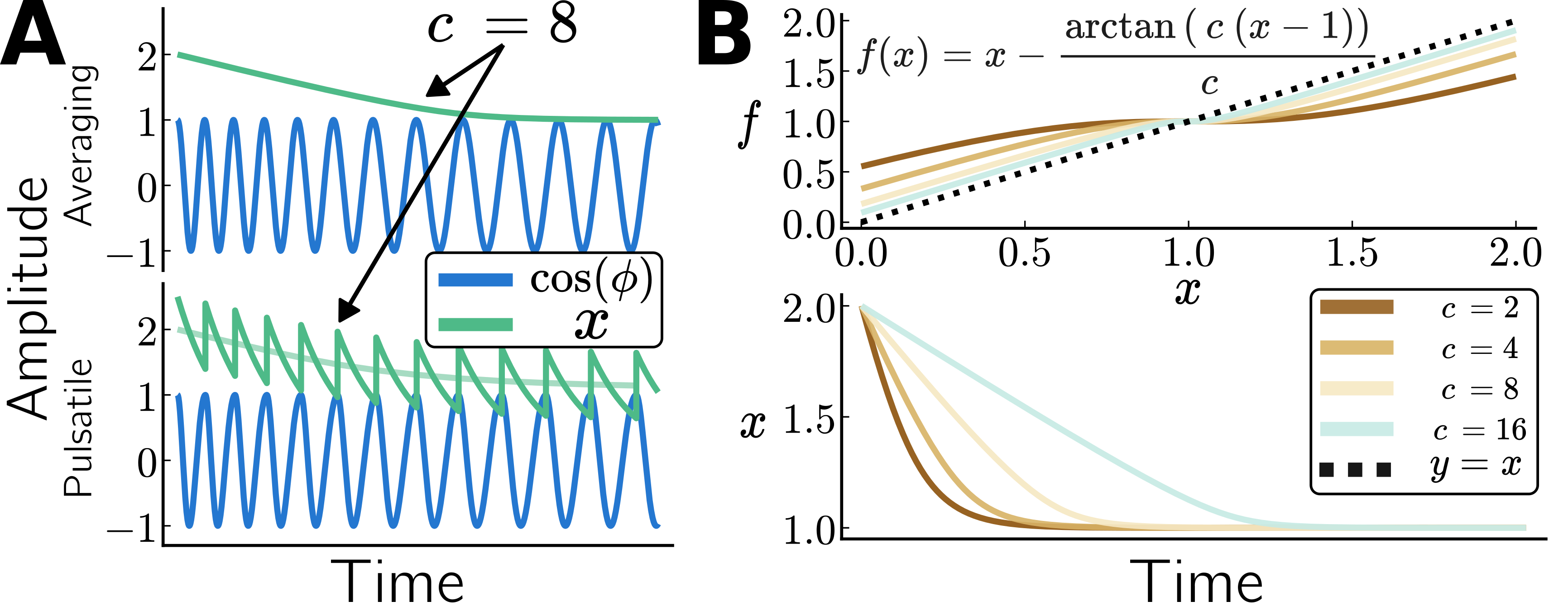}% Here is how to import EPS art
\caption{\label{fig:fig2} Behavior and conditions for models with a memory variable A) The two plots show examples of frequency convergence when starting away from the stable state for the Averaging memory system (top) and the Pulsatile memory system (bottom). $\alpha = 1$. B) Examples of possible coupling functions, with $c_1=c_2=c$ for $f$ and their associated decay rates (for the Averaging-Core Phase system). When $f \sim x$, $x$ decays linearly, while when $f \sim 1$, $x$ decays exponentially. $\alpha = 1$.} 
\end{figure}

We aim to formulate the problem of phase oscillators with memory in the simplest possible way. We consider a generic phase oscillator, called $\phi$, which would correspond to a classical "clock" in the absence of other variables. However, we assume that the frequency of this oscillator called $f$, depends on another variable called $x$, which can be any biological parameter of the system (e.g. some metabolite concentration, a degradation rate, etc.). A natural choice might \textit{a priori} be to assume that $f(x) \sim x$, i.e. that one variable $x$ in the system directly fixes the frequency so that the introduction of an extra step $f$ might seem unnecessary. However, we will see below that in the presence of feedback, one needs an additional non-linearity for the system to be stable, justifying the introduction of such an intermediate.

Indeed, our key additional assumption is that $x(t)$ is some function of previous phases $\phi(u), u\leq t$, thus storing a \textit{memory} of the past of the clock. Since the frequency $f$ is a function of $x$, changes of $x$ possibly lead to frequency changes. Such a model is in particular inspired by what is observed in the segmentation clock. In it, at least two oscillating pathways, Notch and Wnt, were suggested to influence their respective periods and can entrain one another \cite{Sonnen2018, Lauschke2013, francois2024}. This is reminiscent of models of positional information based on the computing of phase difference \cite{Goodwin1969, Beaupeux2016}. Notice that $\phi$ and $x$ are of a different nature: $\phi$ is a phase variable, effectively capturing in an abstract way the state of the system \cite{Kuramoto}, while $x$ is an observable variable of the system such as the concentration of a protein. For this reason, the memory encoded by $x$ is a very different kind of memory than, say, the one in classical delayed oscillators \cite{Lewis2003} where $x$ is a past value of some variable with direct influence on the core mechanism of oscillations. Here, $x$ only modulates oscillations (which arise independently of it), and can potentially be of any nature, as long as it depends on the past values of the clock. It is akin to a memory for two reasons. First, the value of $x$ is constantly \textit{retrieved} through $f$ to influence the frequency of the clock. Second it can be \textit{stored}, and we now introduce two types of couplings from $\phi$ to $x$  , allowing for such storage, conceptualized in Fig. \ref{fig:fig1} C

\subsubsection{Averaging memory}
In the first instance, we consider the following model :
\begin{equation}\label{eq:avg_syst}
\begin{cases}
\dot{\phi} = 2\pi f(x) \\
\dot{x} = \alpha \left(\frac{\dot{\phi}}{2\pi} - x \right).
\end{cases}
\end{equation}

where $\alpha$ sets the timescale of convergence for $x$. We will typically take $\alpha$ small so that $x$ smoothly varies. In such a situation, the memory storage process effectively acts as an averaging of the past frequency, encoded through $x$. We will consequently call this model "averaging memory".

Since $x$, in turn, influences $\phi$, the function $f$ can not be arbitrary. In the absence of any other external signal, one gets for $x$ 
\begin{equation}
\dot x = \alpha(f(x)-x)
\end{equation}
which should have at least one stable fixed point $x_*=f(x_*)$, defining the stable natural intrinsic frequency $\omega_i=2 \pi x_*$. In the following, we rescale time and $x$ so that $x_*=1$. Thus, we also have for the intrinsic frequency $f_i=1$ and the corresponding intrinsic period $\tau_i=1$. Stability further imposes that $f'(x_*)<1$. This means that $f$ has to be sublinear close to the fixed point: for $x>1$, the system slows down to decrease its frequency towards $1$ while for $x<1$ the system speeds up. Figs. \ref{fig:fig2}A,B show the dynamics of such oscillator following a perturbation in $x$ (all simulations presented are available at this GitHub repository, \url{https://github.com/cmdenis/2024-frequency-memory}, see Appendix \ref{app:numerical_integration} for more details).

As said above, a natural choice for $f$ would be $f(x)= x$. The main issue is that we would then get an infinite number of fixed points. Furthermore, we would get $f'(1)=1$ which would then be only marginally stable. To circumvent this, we decided to add a small non-linearity :

%This motivates the following functional form used for simulations presented in this paper

 \begin{equation}
 f(x)=x+\sigma(x)\label{eq:f}
 \end{equation}
 where $\sigma(x)$ is a sigmoid function capturing the non-linearity of $f$ with respect to the internal memory. The choice of a sigmoidal function was motivated by experimental data (see below) and mathematically by the fact that, for bigger $x$, it does not introduce any supplementary (and non-generic) non-linearity that gives rise to new fixed points.

 The conditions $f(1)=1,f'(1)<1$ translate into $\sigma(1)=0,\sigma'(1)<0$. Remarkably, we will show below that multiple aspects of entrainment depend solely on the non-linearity $\sigma(x)= f(x)-x$, further justifying the introduction of such a function. 
 
 We typically consider small $\sigma$ so that $f(x)$ is a small perturbation of the memory variable $x$ controlling the frequency. In the following, we consider functions of the form
 \begin{equation}
 \sigma(x)=-\frac{\arctan c_2(x-1)}{c_1}  \label{eq:sigmoid}
  \end{equation}
 where $c_1$ and $c_2$ are positive constants, as illustrated in Fig. \ref{fig:fig2} B, which ensures that $\sigma'(1)<0$. Big $c_1$ ensures only weak non-linearity. Except otherwise mentioned, in simulations we set $c_1 = c_2 = 8$. Notice also that, for such a choice of $\sigma$ and small and big $x$, $f$ is linear in $x$ up to an additive constant.

%The precise choice we made for the function $f$ will be motivated below when discussing entrainment in light of experiments. 

%In figure \ref{fig:fig2}B, we show how curves continuously approaching either a completely horizontal (non-adaptive) or perfectly linear relationship gives rise to a spectrum of decays going from exponentially shaped decays to linearly shaped decays.

\subsubsection{Pulsatile memory}

The coupling above relies on an implicit averaging of phase dynamics by $x$, which might be difficult to implement biologically because the phase is an abstract variable associated with a limit cycle. We thus propose and study a more mechanistic model for such coupling. It has been recently observed that the coordination of gene expression can be done through pulsatile activations, which provide a natural, and possibly ubiquitous mechanism for frequency sensing by biochemical networks \cite{Cai2008}. The segmentation clock itself seems to present features of pulsatile behaviors \cite{eck2024}, which is also consistent with the idea that it is poised close to a SNIC bifurcation \cite{Jutras-Dube2020}, known to naturally display pulsatile dynamics (type I oscillators \cite{Izhikevich, francois2024}). Motivated by those observations and the model proposed in \cite{Cai2008}, we introduce a "pulsatile memory" model, where pulses in the $x$ variable are generated at a specific phase for $\phi$. Biologically, this could be done for instance, at the maximal value of some oscillatory variable, or if the oscillating system described by $\phi$ is itself pulsatile, this can be done at the pulse itself. We thus study the resulting ODEs:

\begin{equation}\label{eq:pulse_syst}
    \begin{cases}
      \dot{\phi} =& 2\pi f(x) \\
      \dot{x} =& \alpha \left( r \delta \left(\textbf{mod}(\phi, 2\pi)\right) -x\right).  
    \end{cases} 
\end{equation}
 $\delta$ is the Dirac delta function centered around 0, and for convenience, we also define $r\alpha$ as the magnitude of the pulse. To get an intuition for this system, one can think of what happens in the limiting cases. If $\phi$ completes many cycles in a given amount of time, then the pulses begin to "stack up" raising the average value of $x$ since it doesn't get the chance to decrease significantly. Conversely, if very few cycles are completed by $\phi$ in that same amount of time, then $x$ will have the time to decay closer to 0, which will make the average value of $x$ low. All in all, such a mechanism allows $\phi$ to modulate the level of $x$. 

Notice that in the pulsatile memory model, $x$ is oscillating as well, at the same frequency as $\phi$, a situation again reminiscent of somitogenesis with the coupling of multiple biological oscillators \cite{Sonnen2018, Goldbeter2008}. Similar to the averaging memory case, $r$ can not be arbitrary for stability. By integrating Eq. \ref{eq:pulse_syst} over one period with periodic $x$, one gets 

\begin{equation}
0=- \langle x\rangle + \frac{r}{\tau_i}
\end{equation}
where we define $\langle x \rangle=\frac{1}{\tau_i}{\int_0^{\tau_i}} x(t) dt$, the average value of $x$ over one period. Let's take for the stable period $\tau_i \sim 1$ and similar to the averaging memory case $\langle x\rangle=1$. We then get $r = 1$, the choice we make throughout this work \footnote{For $r=1$, the period exactly is $1$ only for $\alpha \rightarrow 0$. The reason is that, when integrating the Dirac function, one needs to be careful about the change of variable $\phi \rightarrow t$, for which the Jacobian is $1$ only if $x$ is almost constant over one cycle, which is the case only when $\alpha\sim 0$. See Fig. \ref{fig:fig10} for other values of $\alpha$}. Other choices are possible for the same $f(x)$ but would then simply change the stable frequency of the oscillator, which can then be put back to $1$ by rescaling time (Appendix \ref{app:var_r}, Fig. \ref{fig:figS2} for different choices of parameters). Given some initial conditions away from the stable state, the system gradually adjusts its frequency to converge back to the stable long-term frequency, at a rate proportional to $\alpha$, as shown in Fig. \ref{fig:fig2}A bottom.

\subsection{Entrainment of the core phase oscillator $\phi$}

Interesting properties occur when those models are perturbed. For a standard limit cycle oscillator, classical entrainment theory reduces to a one-dimensional phase equation reproduced in Appendix \ref{app:classical_entrainment}, Eq. \ref{eq:Hdefinition}. However, with two oscillating variables ($\phi,x$), the system becomes "unclocklike" and new entrainment modalities appear.

We first focus on entrainment on the intrinsic phase $\phi$, which we refer to as the "core" phase oscillator. We introduce an external oscillator $\theta$, with frequency $\omega$, and introduce a coupling function $H$ and a coupling constant $k$ such that the modified equations are :

\begin{equation}
\begin{cases}
\dot{\phi} = 2\pi f(x) + \underbrace{ k\ H(\theta - \phi)}_\text{Core Phase Entrainment} \label{eq:core}\\
\dot{\theta} = \omega \\
%\dot{x} = \alpha \left(\frac{\dot{\phi}}{2\pi} - x \right) 
\end{cases}
\end{equation}
with $x$ dynamics prescribed by either Eq.\ref{eq:avg_syst} or Eq. \ref{eq:pulse_syst} depending on the internal coupling model considered. For convenience, we also define the entrainment period $\tau=\frac{2\pi}{\omega}$.

Fig. \ref{fig:fig3}A illustrates how both Averaging and Pulsatile memory models are entrained by the external oscillator using one of the functions defined in Fig \ref{fig:fig2}B, and a standard Kuramoto coupling $H=\sin$. Of note, the value of $x$ is changing due to entrainment. Since $x$ is both a measure and a control of the internal frequency of the phase oscillator (via the weakly nonlinear function $f(x)=x+\sigma(x)$ with small $\sigma$), it means that the system is adapting its intrinsic frequency in response to the external oscillator.

 Fig. \ref{fig:fig3}B, middle, illustrates the advantage and versatility of systems with frequency adaptation: when ramping up the frequency $\omega$ of the external oscillator $\theta$, the system can dynamically adjust its frequency to follow the external oscillator over a broad range of frequencies. We contrast this behavior with the standard case, where the internal frequency is maintained to $1$ where the system is only entrained for the range of $f$ defined by $2\pi|1-f|<1$ (from Appendix, Eq. \ref{eq:freq_entrained} taking $\omega_i=2\pi$ and $H=\sin$), Fig. \ref{fig:fig3}B, bottom.

\subsection{Arnold Tongues, hysteresis, and maps for Core Phase entrainment}

%\subsubsection{Arnold Tongues and entrained phase}

\begin{figure}
\includegraphics[width=8.5cm]{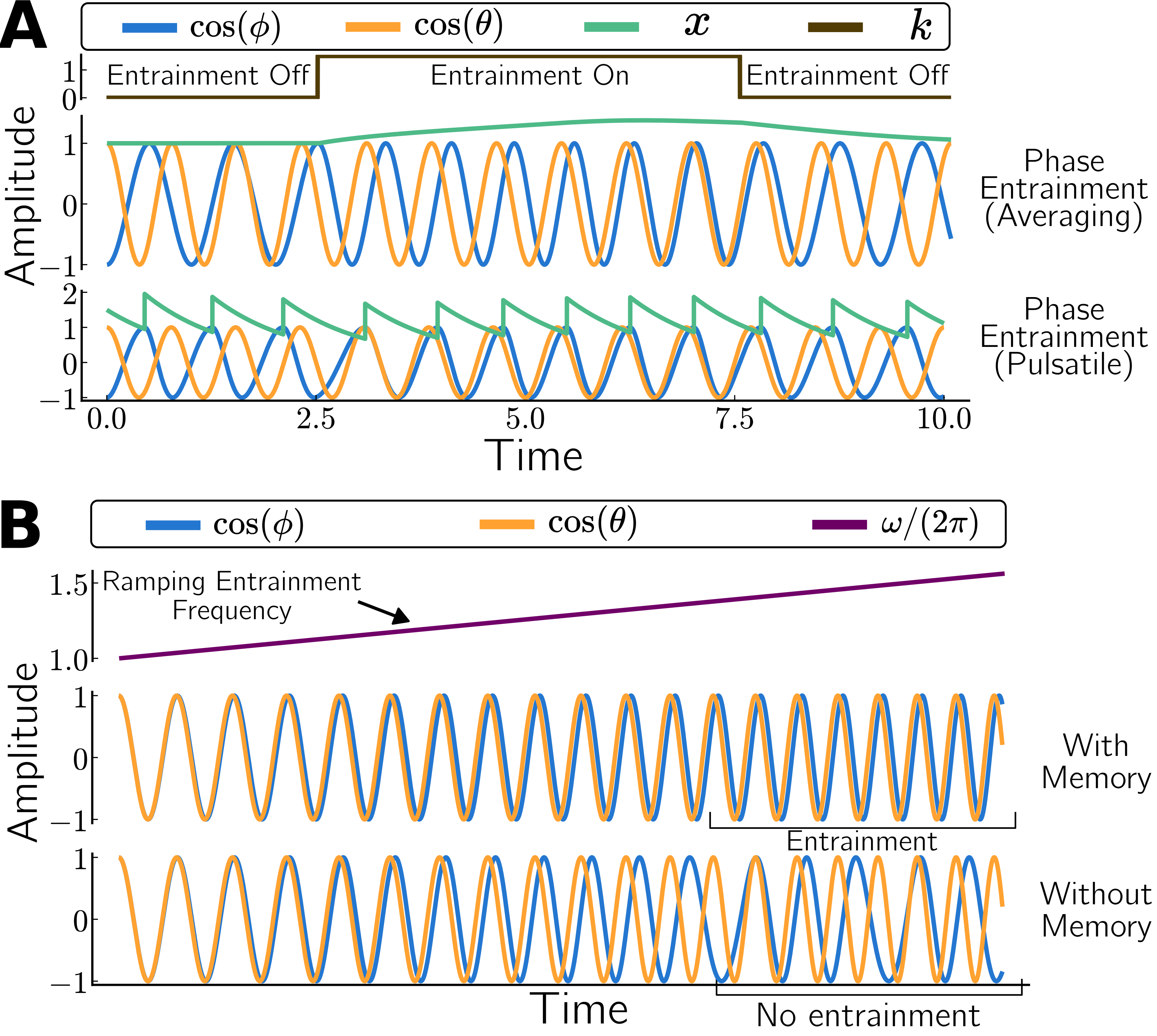}% Here is how to import EPS art
\caption{A) Comparison of different entrainment systems. Each subplot is divided into different regimes. Initially, the system is unentrained. Once the coupling is activated, the two oscillators enter a transient state before eventually becoming synchronized. When the coupling is removed, $\phi$ returns to its stable frequency/period. Middle subplot: Entraining $\phi$ with $\theta$ using the Averaging memory-Core Phase entrainment system. Bottom subplot: Entraining $\phi$ with $\theta$ using the Pulsatile memory Core Phase entrainment system. $\omega/(2\pi) = 1.3$, $\alpha=1$. B) Comparison between a system with memory and a system with no memory, if entrained by an external oscillator whose frequency is ramped up, with $k=1.5$, $\alpha = 5$.} \label{fig:fig3}
\end{figure}

\begin{figure}
\includegraphics[width=8.5cm]{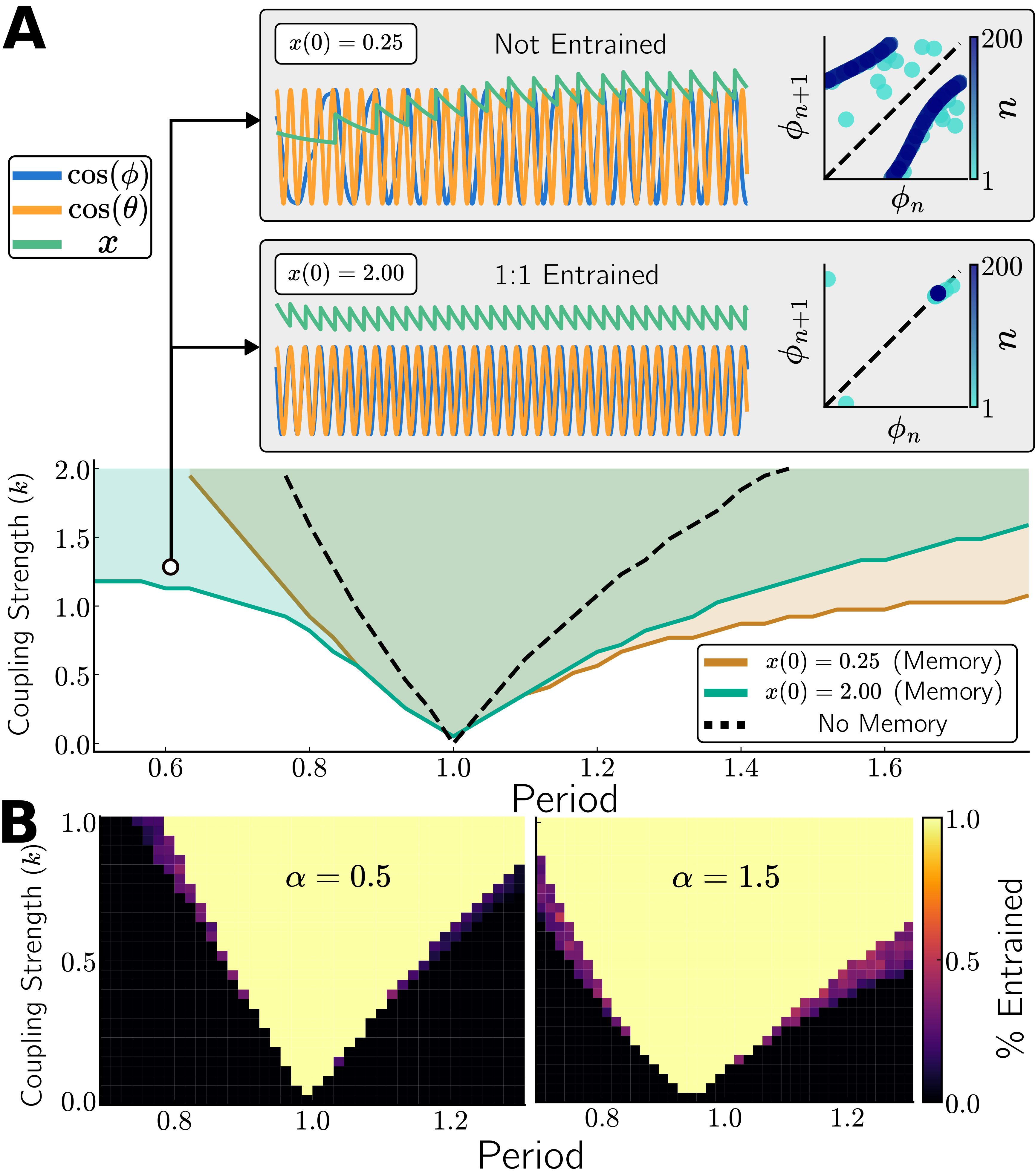}% Here is how to import EPS art
\caption{A) Arnold tongues for the Pulsatile memory-Core Phase entrainment system for different initial conditions, contrasted with a system with no memory. The system with memory has wider tongues. The Arnold tongues were computed numerically by checking the convergence of phase $\phi$ using stroboscopic maps over 200 cycles (top boxes). We show sample runs and the effect initial conditions have on entrainment. The stroboscopic maps for the systems are also shown next to their corresponding trajectory. Entrained (1:1) trajectories have the points in the map converge towards a single value. The color gradient indicates the time, in cycles, at which the Poincare section was taken in the simulation. $\alpha = 0.5$. B) Probabilistic Arnold Tongue for the Pulsatile memory-Core Phase entrainment system with $\alpha=0.5$ and $\alpha = 1.5$. Varying the $\alpha$ parameter affects the width of the tongue. $x(0)=1.1$.} \label{fig:fig4}
\end{figure}

Fig. \ref{fig:fig4} displays the entrainment range of oscillators as a function of the rescaled period $2\pi/\omega$ and of the coupling strength $k$ for $H=\sin$, generally called "Arnold tongues" \cite{glass2001}, displayed here for the pulsatile memory system, (see Appendix \ref{app:tongue-average}, Fig. \ref{fig:figS3} for equivalent tongues with averaging memory), for different initial conditions of the system. Fig. \ref{fig:fig4}A illustrates the range of possible entrainment for various values of initial $x$ (a set of parameters is labeled as entrained if numerically a $\phi(t=0)$ is found that makes the trajectory entrained), while Fig. \ref{fig:fig4}B shows what happens for a fixed initial value of $x(0)=1.1$ as we vary the initial condition for the phase $\phi(t=0)$ and $\alpha$. A comparison of the tongues with a fixed $f=1$ (no memory) is shown in Fig. \ref{fig:fig4}A.

Focusing first on Fig. \ref{fig:fig4}A, the tongue for the system with memory widens for higher coupling strength compared to the equivalent one without memory, meaning that the system is much easier to entrain as already illustrated in Fig. \ref{fig:fig3}. This is reminiscent of the large tongues obtained for the segmentation clock in \cite{Sanchez2022}. However, we further observe that the initial conditions for $x$ strongly matter to have entrainment within a given tongue, a new "unclocklike" property of such systems. Simulations that started with lower initial $x$ can usually get entrained to longer periods, and simulations that started with higher initial $x$ can get entrained to shorter periods, Fig. \ref{fig:fig4}A. 

Conversely, one can keep $x$ constant and change the initial phase of $\phi$, to observe probabilistic Arnold tongues, Fig. \ref{fig:fig4}B. The Arnold tongues are typically wider for bigger $\alpha$ because $x$ varies much more and relaxes more rapidly in such cases. However, close to the boundaries of the entrainment range, some initial phase conditions do not allow for entrainment because $x$ is pushed at too low or too high values, an effect amplified for big $\alpha$. Again, such initial condition dependencies are not possible in classical "clocks" in phase response theory, because entrainment is associated with a fixed point equation on the phase, which either has a solution or not.

\begin{figure}
\includegraphics[width=8.5cm]{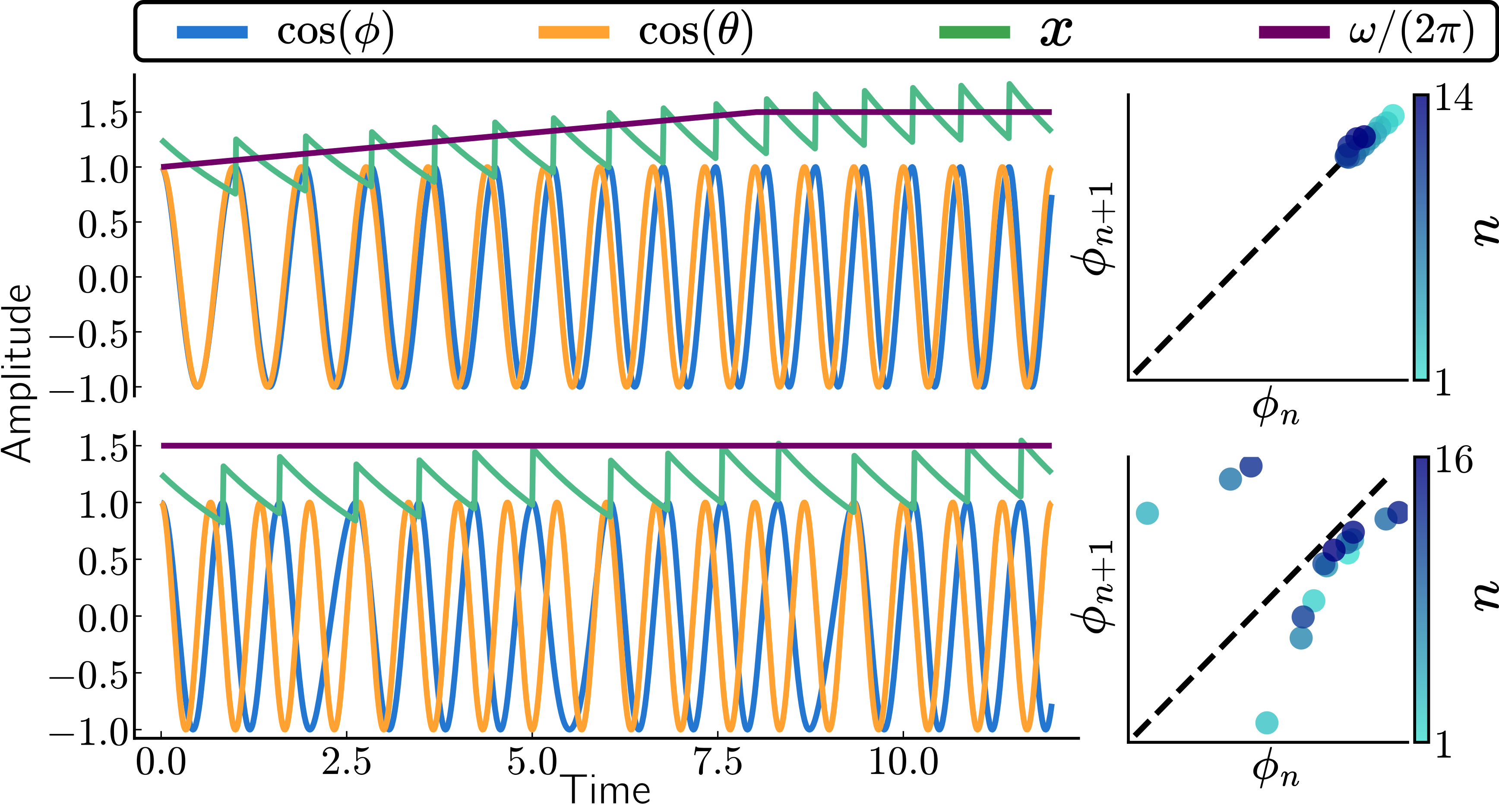}% Here is how to import EPS art
\caption{ Experimental scheme to test for hysteresis. Parameters are $k=1.8$, $\alpha = 0.5$. (Top) For the Pulsatile memory-Core Phase entrainment system, the entrainment frequency ($\omega/(2\pi)$) is started at 1 and slowly ramped up to 1.5: entrainment persists, as indicated by the fixed point in the stroboscopic map. The color gradient indicates the time, in cycles, at which the Poincare section was taken. (Bottom) The system's entrainment frequency ($\omega/(2\pi)$) is started immediately at 1.5: no entrainment, as indicated by the lack of a fixed point in the stroboscopic map.} \label{fig:fig5}
\end{figure}

This is the first type of hysteresis, where the entrained/not entrained state depends on the history of the system via $x$, further illustrated in Fig. \ref{fig:fig5}. We show two different protocols: in the top of Fig. \ref{fig:fig5}, the angular velocity of the external oscillations ($\omega$) is started at the natural intrinsic value corresponding to $f=x=1$, and then slowly increases by $50 \%$ while keeping the system entrained. However, if this angular velocity is suddenly changed from the same initial condition, the system does not entrain, Fig. \ref{fig:fig5} bottom. In Appendix \ref{app:hysteresis}, Fig. \ref{fig:figS4} shows similar behavior for the Averaging memory-Core Phase entrainment system when ramping the entrainment frequency at different rates.

%This is true for both the pulsatile system and the direct frequency coupling system. This is a major difference from usual non-adaptive systems.
%As said above, the addition of the new adaptive variable $x$ introduces new features absent from classical entrainment theory of single phase variable.

 %In classical Arnold tongue theory, one finds a well-defined cutoff point marking the boundary of a given tongue (dashed lines in figure \ref{fig:fig3}A).  Different tongues can overlap so that multistability can be observed depending on the initial conditions of $\phi$, but one does not observe cases where the system is entrained for some initial conditions and not entrained for other ones. 

%In the adaptive system we study here, the addition of the variable $x$ complexifies this picture, since, as said above, the range of entrainments also depends on the initial values of $x$.

%This is further illustrated in Fig. \ref{fig:fig5} A : with the same external oscillator frequency, depending on the initial conditions in $x$, one can entrain the oscillator or not. 

\begin{figure}
\includegraphics[width=8.5cm]{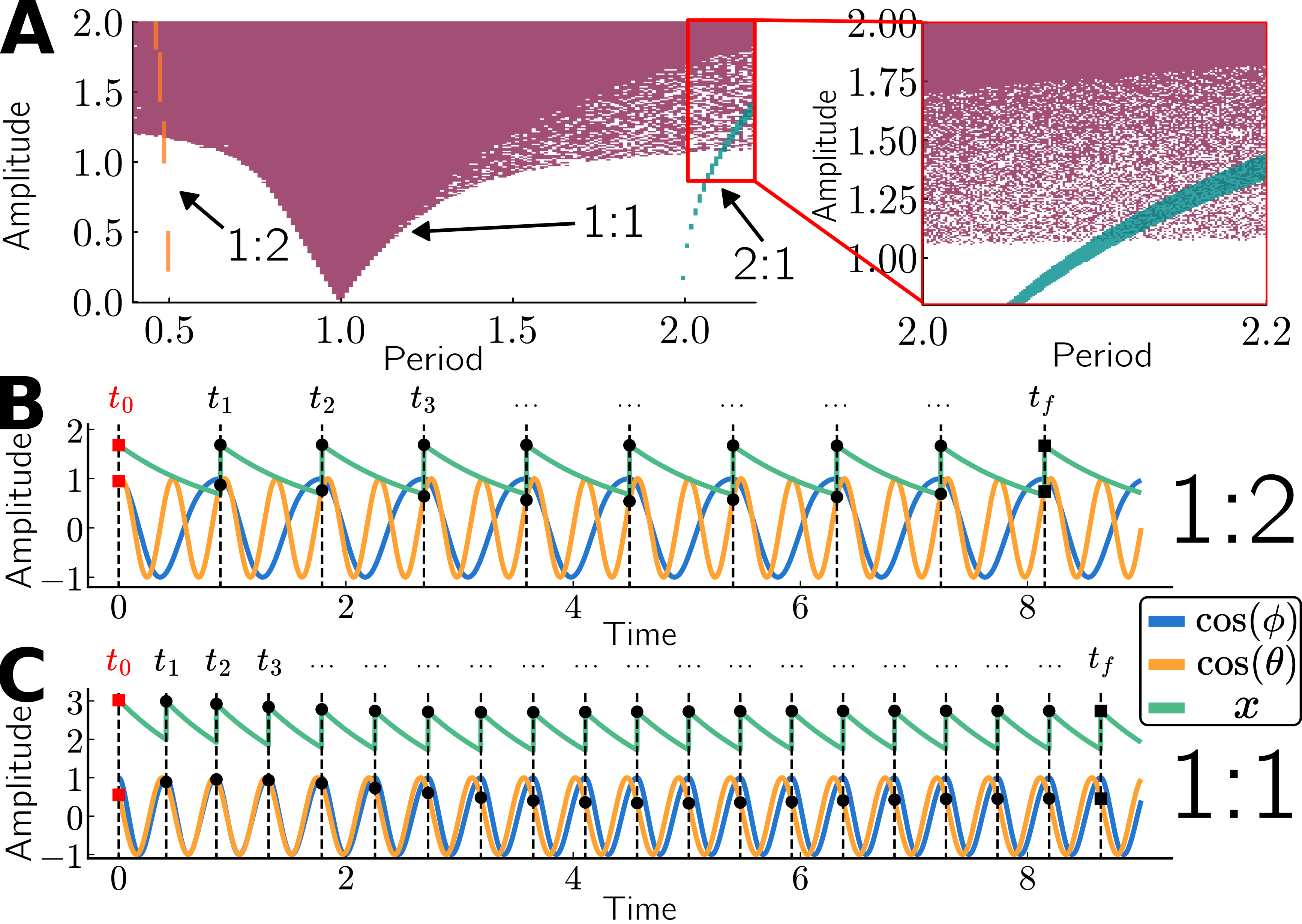}% Here is how to import EPS art
\caption{A) Arnold tongues for the pulsatile memory-core phase entrainment system. We show the 2:1, 1:1, and 1:2 tongues. For each point, the system is integrated for randomized initial conditions (in $x$ and $\phi$) and then checked for entrainment. We observe the 2:1 and the 1:2 tongues superimposed on the "speckled" 1:1 tongue, showing bistability. B) Example of the entraining process for the 1:2 regime. The dots at each point correspond to the trajectories plotted in Figure \ref{fig:fig7}B and C. Notice the convergence of the dots to the fixed point. $\omega/(2\pi)=2.2$. $k = 1.5$. $\alpha = 1.0$.} \label{fig:fig6}
\end{figure}

\begin{figure}
\includegraphics[width=8.5cm]{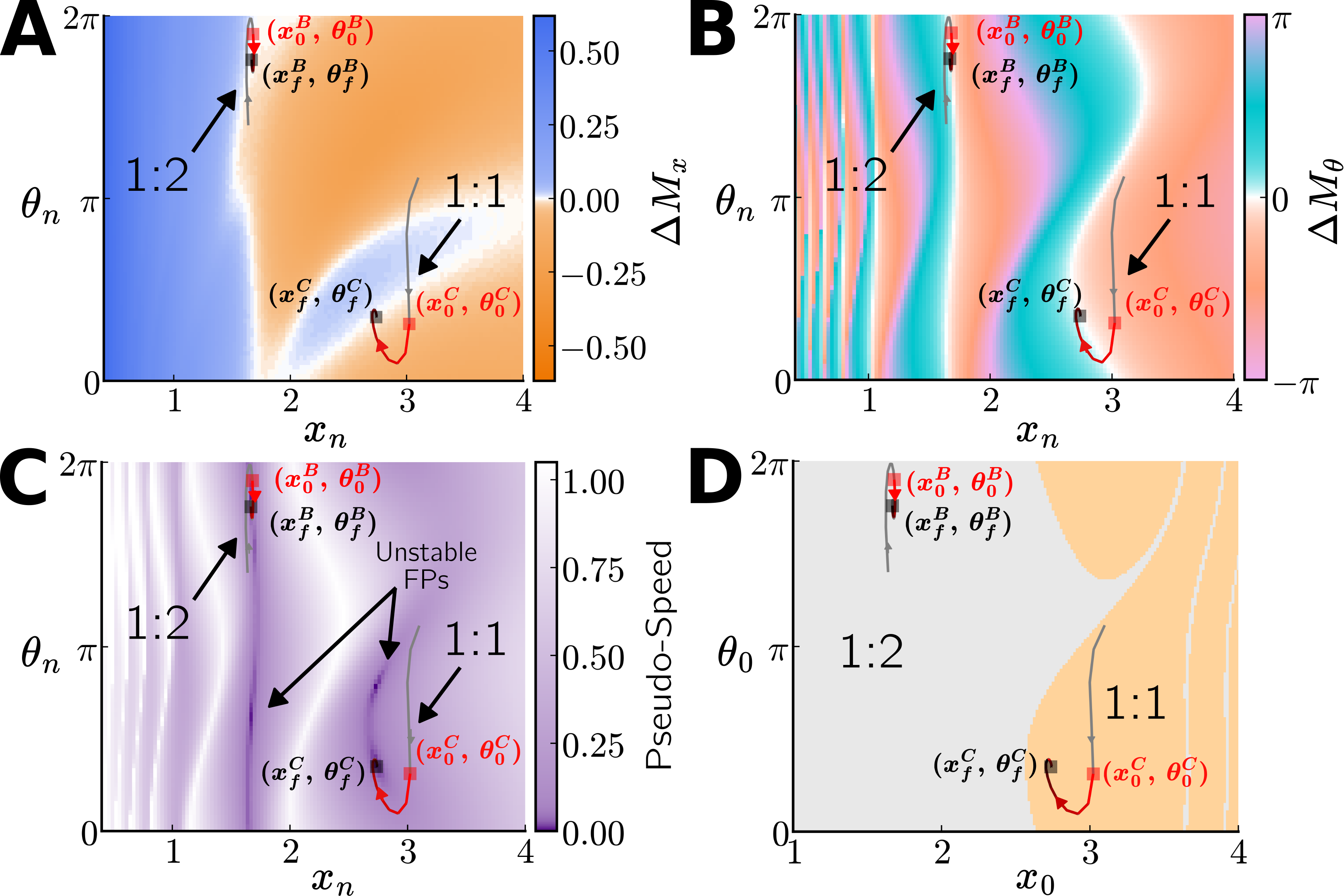}% Here is how to import EPS art
\caption{A) Map $\Delta M_x$ showing the difference in $x$ between each consecutive pulse for the pulsatile memory-core phase entrainment system. Lines going through the center of white regions form nullclines, which allows us to identify fixed points. We see examples of trajectories, with their initial $(x_0^{B/C},\ \theta_0^{B/C})$ and final $(x_f^{B/C},\ \theta_f^{B/C})$ states, attracted to the fixed points having 1:1 and 1:2 entrainment regimes. A gray line extends this further "back in time" to illustrate more clearly the attractive nature of the fixed point. The $B$ and $C$ labels link those trajectories to those presented in figure \ref{fig:fig6}B,C. B)  Map $\Delta M_\theta$ showing the difference in the phase of $\theta$ between two consecutive pulses, with examples of trajectories. $\omega/(2\pi) = 2.2$. $k=1.5$. $\alpha = 1.0$. C) Normalized "pseudo speed" of the mapping in A and B. We see the 4 fixed points of the system, 2 of which are stable and 2 unstable. Example trajectories displayed in fig \ref{fig:fig6}B,C are plotted and are attracted to the stable fixed points. D) Basin of attraction for the pulsatile memory-core phase coupling. We see the two fixed points corresponding to the 1:1 and 1:2 entrainment regimes and example trajectories already presented in Fig. \ref{fig:fig6}. The basin seems continuous and features an elongated ribbon-like region extending for higher initial $x$, resulting in 1:1 entrainment. Note the periodicity, and continuity of the vertical axis.}
 \label{fig:fig7}\end{figure}

% Another form of hysteresis is observed for the pulsatile system, with a new kind of bistability compared to classical entrainment theory, illustrated in Fig. \ref{fig:fig6}A and B. Classically, a $1:1$ tongue is observed if the frequency of the external oscillator $\theta$ is close to the intrinsic frequency of the oscillator $\phi$. However, if the $x$ variable is perturbed,  new entrainment regimes appear as the intrinsic frequency of the oscillator changes because of the feedback. For instance, if $x$ is initialized at a low value, this poises the oscillator toward low frequency.
%

 Fig. \ref{fig:fig6}A further illustrates the entrainment regimes reached through randomized initial conditions. A speckled 1:1 tongue far away from the stable intrinsic frequency is the hallmark of the bistability mentioned above (Fig. \ref{fig:fig5}), where the system can be either entrained or not entrained depending on initial conditions. We also observe 2:1 and 1:2 tongues that are superimposed unto the 1:1. This demonstrates another type of bistability, on entrainment itself, i.e. for the same parameters, the system produces different entrainment outcomes, depending on the initial conditions, thus giving rise to a second type of hysteresis. Figs. \ref{fig:fig6} B and C illustrate what happens for an entrainment frequency that is close to twice the intrinsic one: either the system keeps its intrinsic frequency of $\sim$1, so that it is entrained with two cycles of the external oscillator for one intrinsic cycle, Fig. \ref{fig:fig6}B) or the system speeds up its internal frequency (via higher values of x) to match the external frequency and remain in the 1:1 tongue, Fig. \ref{fig:fig6}C

%We also define two other maps for the separate variables: 1) $M_x$, such that $M_x(x_n, \theta_n) = x_{n+1}$ and 2) $M_y$, such that $M_\theta(x_n, \theta_n) = \theta_{n+1}$.

 This behavior can be understood in terms of attractors and fixed points of return maps, as shown in Fig. \ref{fig:fig7}. Given a set of parameters $k,\omega, \alpha$, it is possible to (numerically) integrate the differential equations to define the following bidimensional map:
 
\begin{equation}
 (x_{n+1},\ \theta_{n+1})=M(x_n,\ \theta_n) 
 \end{equation}
where $x_n$ and $\theta_n$ are the respective values of $x$ and of the entrainment signal $\theta$ right after the pulse in $x$ occurring at $t_n$ (where $\phi \mod 2\pi = 0$). See Figs \ref{fig:fig6} B-C for visualization of the pulses. Importantly, since at $t_n$ $\phi=0, x=x_n, \theta=\theta_n$, this system is fully closed so that the map can be exactly derived, and, consequently, the entire future behavior of the system is determined by such maps.

 To better visualize this map, we plot the respective values of $\Delta M_{x}=x_{n+1}-x_n$ and of $ \Delta M_{\theta}=\theta_{n+1}-\theta_n$ as a function of $(x_n,\theta_n)$ in Figs. \ref{fig:fig7} A,B. Respective zeros for $\Delta M_{x}$, and $ \Delta M_{\theta}$ correspond to white lines in Figs. \ref{fig:fig7} A-B. Their intersections define fixed points in $(x,\theta)$, corresponding to possible entrained states. To visualize their location, we also plot a pseudo velocity scalar field $v=\sqrt{ \Delta M_{x}^2+\Delta M_{\theta}^2}$, Fig. \ref{fig:fig7}C, so that zero velocity corresponds to points where both $\Delta M_{x}=\Delta M_{\theta}=0$.
 
 We call $(x_f,\theta_f)$ the respective values of $x,\theta$ at the fixed points, and index them with the corresponding panel of Fig. \ref{fig:fig6} B-C. We also draw the trajectories corresponding to Fig. \ref{fig:fig6} B-C on each panel of Fig. \ref{fig:fig7}. For parameters used in Fig. \ref{fig:fig7}, we see on panel C that there are 4 fixed points. The stability of those points can further be inferred by looking at the local sign of $ \Delta M_{x},  \Delta M_{\theta}$: e.g., the "blue" regions in $x$ correspond to positive $\Delta M_{x}$ and thus move towards higher $x$. Therefore, a fixed point is stable in $x$ only if it is at the intersection of a positive-to-negative (blue-to-orange) transition in $ \Delta M_{x}$ for increasing $x$ (and there are similar rules for $\theta$).

 We thus see from Figs. \ref{fig:fig7} A-C that there are 2 stable and 2 unstable fixed points. Each stable fixed point corresponds to one of the tongues: the 1:2 tongue corresponds to the fixed point $(x^B_f,\theta^B_f)$, with $1<x_f<2$, so with little modification of the intrinsic frequency, while the 1:1 tongue corresponds to the fixed point $(x^C_f,\theta^C_f)$, where $x_f\sim 3$ indicating a big change of the intrinsic frequency. We also notice that the trajectories close to the fixed points tend to follow lines of slower velocity, corresponding to $\theta_{n+1}\sim \theta_n$, Fig. \ref{fig:fig7}B-C. The basins of attraction for those fixed points can also be numerically computed, as seen in Fig. \ref{fig:fig7} D, in particular confirming the importance of the initial condition in $x$: higher values tend to go to the 1:1 fixed point, consistent with the observations made in Fig. \ref{fig:fig4} A. We notice that the complete knowledge of those return maps excludes the existence of other stable fixed points for the range of $x$ considered \footnote{However, one could potentially observe other fixed points for $N:M$ entrainment, for integer $N\neq 1$ by iterating the map, a problem that is considerably increasing in complexity given the bidimensionality of the initial map}. Finally, it is important to point out that such bistability does not exist for traditional systems without memories because the return map is unidimensional, i.e. $\theta_{n+1}=M(\theta_n)$.

Beyond the study of those maps, we can get some intuition on the origin of the bistability. In Supplement Fig. \ref{fig:figS5}, we contrast those maps with pulsatile models without memory, i.e. where $f(x)=1$, so that $x$ is completely slaved to $\phi$. We illustrate what happens for an entrainment frequency close to $1$, for which there is one fixed point both with and without memory Fig\ref{fig:figS5} A-D, and for a double frequency comparable to Fig. \ref{fig:fig6}, for which there is no fixed point without memory, Fig\ref{fig:figS5} E-F. The biggest change induced by the presence of the memory is on the $\theta$ map: in the absence of memory, $\theta_{n+1}$ is a pure function of $\theta_n$ so that the corresponding map $\Delta M_{\theta}=\theta_{n+1}-\theta_n$ has a much simpler structure, with horizontal lines. By contrast, the presence of the memory strongly modifies this map, making $\theta_n$ extremely sensitive to $x$, as visible from the more vertical lines in Fig. \ref{fig:fig7} B and Fig. \ref{fig:figS5} B. This is very intuitive: $x$ directly influences the internal frequency, and thus strongly influences both the timing of its next pulse $t_{n+1}$ and the corresponding value of the phase of the entraining signal $\theta_{n+1}=\theta(t_{n+1})$. As for the $x$ map, the structure of the map is simpler, but we see that memory favors the emergence of more "horizontal" structures for the line $\Delta M_x=0$, Figs. \ref{fig:fig7}A and \ref{fig:figS5} C. This allows for extra crossings with the vertical lines $\Delta M_\theta=0$, resulting in the emergence of new attractors.

%The $x_{n+1}-x_n$ also has a much simpler, monotonic structure : inde

%Those more 'vertical' lines allow for more possibilities of intersections with $$

%$x$ is then completely determined by $\phi$. So both maps in Fig. \ref{fig:fig6} B,C would reduce to one vertical line; futhermore the map in Fig. \ref{fig:fig6} B would be identically $0$, while the value of $\theta_{n+1}$ would then be a pure function of $\theta_n$, corresponding to one vertical line in Fig. \ref{fig:fig6} C. The internal memory thus causes two effects : first the return map are bidimensional, and second the return map for $x$ is non zero. Intuitively, this comes from the fact that, indeed, $x$ adjusts its value in response to the entrainment, until it reaches new stable values, corresponding to white lines in Fig. Fig. \ref{fig:fig6} B. In turns, those two effects creates a new fixed point corresponding to the 1:1 tongue.

%We have shown the maps corresponding to the value of $x$ and $\theta$ at each pulse. It is clear from the example trajectories that this map yields bistability in the 1:1 and 1:2 entrainment outcomes. Given a certain initial condition $(x_0,\ \theta_0)$, what values will eventually result in 1:1 or 1:2 entrainment? We have probed numerically for the initial conditions in $x$ and $\theta$ that result in those two outcomes, as illustrated in figure \ref{fig}B. 

%Here, importantly, the system displays bistability in its intrinsic frequency $f(x)$ due to the feedback between $\phi$ and $x$, which is impossible to observe in standard models where $f$ is constant.

\subsection{Phase of entrainment}

\begin{figure}
\includegraphics[width=8.5cm]{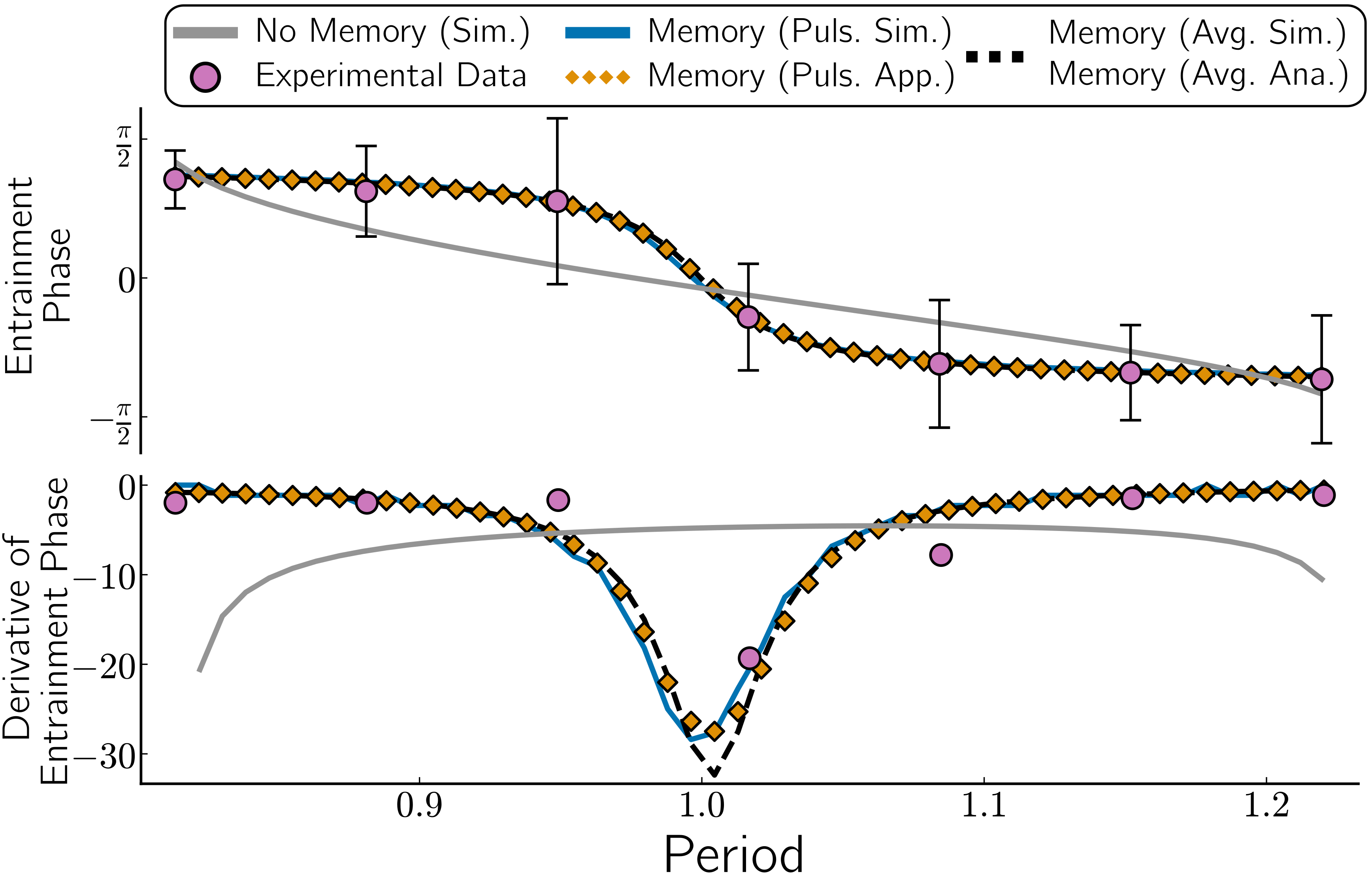}% Here is how to import EPS art
\caption{(Top) Entrainment phase for models with memory (core phase entrainment) and without memory, fitted to the experimental data. The models with memory all present the same overlapping shape. The black dotted curve is the analytically computed phase difference Eq. \ref{eq:entrain_phase_avg} (indistinguishable from the equivalent simulated results), which perfectly overlaps with the simulated pulsatile memory-core phase entrainment system (blue) and Eq. \ref{eq:phi_pulse} (diamond yellow). This contrasts with the $\arcsin$ shape of the system without memory, Eq. \ref{eq:phase_entrained}. For comparison with experiments, we shifted the (arbitrary) phase definition of experimental data for $\theta$ so that $\Delta \phi=0$ for a rescaled period of $1$. Parameters for memory system: $c_1 = 7.6$, $c_2 = 53.6$, $k=1.35$, $\alpha = 0.03$, $\tau_{\text{scale}}=147.5$. Parameters for the system without memory: $k = 1.3$, $\tau_{\text{scale}}=144$. (Bottom) Derivative of the entrainment phase for the system with memory, without memory, and the experimental data, where slight differences between simulations and models are more visible.} \label{fig:fig8}
\end{figure}
We now go back to the 1:1 map, assuming $\omega$ close to the intrinsic angular velocity.
 An important observable associated with entrainment is the stationary phase difference $\Delta \phi= \theta-\phi$ between entrained and entraining oscillators, Fig. \ref{fig:fig8}. Importantly, such phase differences are much easier to measure experimentally with low uncertainty compared to other variables or Arnold Tongue boundaries \cite{Sanchez2022}, and as such constitute one of the best ways to discriminate between different models.

%The phase difference curve also gives us a good intuition into what parameters yield entrainment, as the domain of the curve corresponds to the range of frequencies producing entrainment. 

%Oscillators with frequency adaptation present very different properties, which we illustrate now. 

\subsubsection{Averaging memory}
Studying first the averaging memory model, and assuming that at entrainment $x=x_c$ is approximately constant (which is approximately the case if $\alpha$ is small), and that $\Delta \phi=\theta - \phi$ is constant meaning that $\dot \phi = \omega$, we get : 

\begin{equation}
\omega = 2\pi f(x_c) + kH\left(\Delta \phi \right) \label{eq:omega_entrained}
\end{equation}
It also follows from Eq. \ref{eq:avg_syst} at stationarity that 
\begin{equation}
x_c=\frac{\dot \phi}{2\pi} = \frac{\omega}{2\pi}=\frac{1}{\tau} \label{eq:xc}
\end{equation}

 so that $x$ is indeed a direct measure of the external frequency, as expected by construction.

We can then invert Eq. \ref{eq:omega_entrained} to get :

\begin{equation}
\Delta \phi = H^{-1} \left(\frac{-2\pi\sigma(\frac{\omega}{2\pi})}{k}\right)\label{eq:entrain_phase_avg}
\end{equation}
where $\sigma(x)=f(x)-x$ measures the deviation between $f$ and a pure linear behavior. Notice that, remarkably, the entrainment phase thus provides a direct measurement of the non-linearity $\sigma(x)$ (assuming $H$ is known).

In experiments on the segmentation clock \cite{Sanchez2022}, data fitting suggests that the change of internal frequency with detuning $df/d\omega$ is approximately $1$ for high and low periods. This means that $\sigma$ converges to constant values for high and low $x$, thus justifying the sigmoid form defined in Eq. \ref{eq:sigmoid}. From Eq. \ref{eq:entrain_phase_avg}, we expect convergence of the entrained phase towards constant values at low and high periods. Those values are free parameters of the model but can be fitted so that simulations are in full agreement with the experimental data reproduced in Fig. \ref{fig:fig8}, using $H=\sin$. This result is to be contrasted with the classical entrainment that predicts infinite derivatives at the boundaries of the entrainment ranges in such regimes (Fig. \ref{fig:fig8}, gray line, and Appendix \ref{app:classical_entrainment}). One could be concerned that such plateauing would come from the perfectly linear asymptotic behavior of $f$, in Appendix \ref{app:asymptotic}, Fig. \ref{fig:figS6} we illustrate what happens for different asymptotic slopes, and we still observe sigmoid shapes consistent with experiments for a broad range of periods. 

\subsubsection{Pulsatile memory}

For the pulsatile memory system, similar analytical results can be obtained, see details in Appendix \ref{app:pulse_entrainment}. Assuming the oscillator is entrained, the (entrained) period of the $\phi$ oscillator then matches the period $\tau$ of the entraining signal $\theta$. Recalling that $f(x)=x+\sigma(x)$, by integrating over one entrained cycle from $t=0$ to $t=\tau$, we get from the $\phi$ equation in Eq. \ref{eq:core} 

\begin{equation}
    \frac{1}\tau\int_0^\tau \dot\phi dt=\frac{2\pi}{\tau}=\omega=2\pi\langle x\rangle + 2\pi\langle \sigma(x)\rangle  + k\langle H(\Delta \phi)\rangle. \label{eq:highalpha}
\end{equation}

Between two successive pulses, the dynamics of $x$ simplify into the linear equation $\dot x=-\alpha x$ and can be fully integrated. We take the origin of time $t=0$ right after the pulse of magnitude $r\alpha$, so that $x(t)=x(0) e^{-\alpha t}$ and to get the initial condition, we assume there is entrainment to relate the value of $x$ before and after the pulse, i.e. $r\alpha+ x(\tau)=x(0)$ so that we get

\begin{equation}
    x(t) = r\alpha \frac{e^{-\alpha t}}{1-e^{-\alpha \tau}}.\label{eq:x}
\end{equation}
This equation holds independently of the value of $\alpha$. We further see that $\int_0^\tau x(t) dt=\frac{r\alpha}{\alpha}=r=1$. In the limit of small $\alpha$ so that $\langle x\rangle=\frac{1}{\tau}=\omega/(2\pi)$, showing that, due to entrainment, $x$ adjusts to match the entrainment frequency on average, similar to the averaging memory case (Eq. \ref{eq:xc}). However, $x$ is not constant in that case and thus varies around that average value.

Injecting into Eq. \ref{eq:highalpha}, the $\omega$ terms cancel out and we get 
\begin{equation}
   0= \langle \sigma(x)\rangle  + \frac{k}{2\pi}\langle H(\Delta \phi)\rangle. \label{eq:sigma}
\end{equation}

Remarkably again, this equation relates intuitively and compactly the entrained phase difference between the two oscillators $\Delta\phi$ to the non-linearity of the internal frequency $f$, which is entirely captured by $\sigma(x)=f(x)-x$. Notice that Eq. \ref{eq:entrain_phase_avg} for averaging memory is a mean-field equivalent of Eq.\ref{eq:sigma}.

Further assuming now that $\Delta \phi$ (but not necessarily $\sigma$) is approximately constant over one cycle (see Appendix \ref{app:pulse_entrainment}), we can get a similar form as Eq. \ref{eq:entrain_phase_avg}

\begin{equation}
\Delta \phi = H^{-1} \left(\frac{\left(\omega - \chi(\omega) \right)}{k}\right) \label{eq:phi_pulse}
\end{equation}

with $\chi(\omega)$ given by Eq. \ref{eq:chi}. In Appendix \ref{app:pulse_entrainment} we show that, for our choices of $f$, $d\chi(\omega)/d\omega \sim 1 $ (Eq. \ref{eq:asymp_chi}) for high and low detuning, ensuring that in the pulsatile memory case, $\Delta \phi$ plateaus in the same way as the averaging memory case for high and low detuning.

In fact, assuming that $\alpha$ is small, we can see from Eq. \ref{eq:x} that $x$ is almost constant. At order $0$ in $\alpha$, we get back that $x\sim \frac{1}{\tau}$, and from Eq.\ref{eq:sigma}

\begin{equation}
 H(\Delta \phi)=-\frac{2\pi}{k} \sigma(x)\label{eq:sigma_small_alpha}
\end{equation}

which consistently turns out to be identical to the averaging memory model result in Eq. \ref{eq:entrain_phase_avg}. Thus in the limit of small $\alpha$, there is no difference between the pulsatile and averaging memory models for $\Delta \phi$ at steady state, suggesting that the computation of such phase difference is a good way to show the existence of an internal frequency memory irrespective of the model's details.

Fig. \ref{fig:fig8} further shows the numerical and analytical phase curves for the two types of coupling and are further compared to a standard system without memory, and the experimental data presented in Sanchez et al\cite{Sanchez2022}.

\subsection{Frequency Entrainment }\label{sec:entrain_memory}

\begin{figure}
\includegraphics[width=8.5cm]{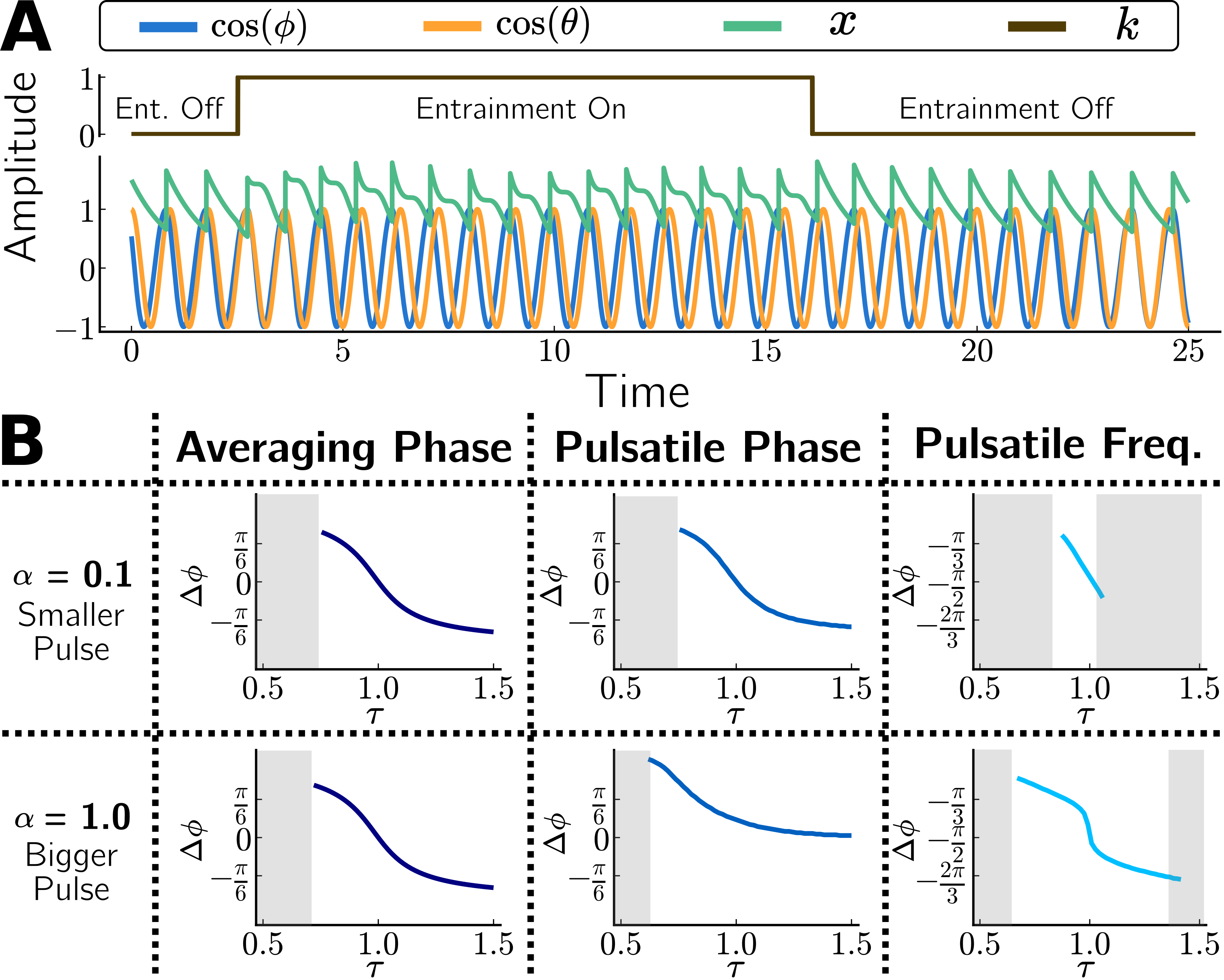}% Here is how to import EPS art
\caption{\label{fig:fig9}A) Example of entrainment for the pulsatile-frequency system. The system is initially unentrained until the coupling is activated, after which it becomes entrained. The coupling is stopped afterward. The timescale of the entrainment is slower for the frequency entrainment than for the core phase entrainment. Notice the deformation of the $x$ curve by the sinusoidal coupling. $k=1.0$, $\alpha = 1.0$, $\omega /(2\pi)=1.1$. B) Comparison of entrainment phases for the different models under consideration, with their entrainment ranges. The gray regions correspond to no entrainment. Simulations were all initialized with $x(0)=1$. Other examples of such curves for different couplings are qualitatively identical or similar and shown in Appendix \ref{app:var_coup_freq}. $k=1.5$.} 
\end{figure}

It is natural to investigate entrainment performed on $x$, the memory variable (instead of $\phi$). If $f$ is weakly non-linear, we expect $x$ to be more directly related to the frequency, so we call this "frequency entrainment", but, for this reason, it is not clear that $x$ has enough phase dependency to allow for entrainment (defined as a fixed point for the phase in a stroboscopic map). Indeed, we could not entrain the averaging memory model, probably because of the explicit averaging performed by $x$ which erases any phase information. 

In the pulsatile memory system, $x$ is oscillating and thus "encodes" some phase information. We study
\begin{equation}
    \begin{cases}
      \dot{\phi} =& 2\pi f(x)  \\
      \dot{x} =& \alpha\left( \underbrace{ r \delta \left(\textbf{mod}(\phi, 2\pi)\right)}_\text{Pulses} - x \right)   + \underbrace{k \sin\left(\theta\right)x}_\text{Freq. Entrainment},
    \end{cases}\label{eq:freq_coupling}
\end{equation}

which allows for an explicit analytic expression of $x$ if there is entrainment (see Appendix \ref{app:small_alpha_delta_phi} and Eq. \ref{eq:ana_x_freq}).

Entrainment on this system is illustrated in Fig. \ref{fig:fig9}A and Fig. \ref{fig:fig9}B further compares various entrainment modalities. For small $\alpha=0.1$ we observe a much smaller entrainment range for frequency entrainment compared to the phase entrainment, and an associated smaller range of possible $\Delta \phi$. This is consistent with the fact that $x$ indeed  erases much of the phase information: it is thus difficult to entrain, and the system must be adjusted to a specific phase to do so.

To understand what happens, it is again useful to integrate Eq. \ref{eq:freq_coupling} over one cycle right after the pulse in $x$ to get 

\begin{eqnarray}
\int_0^{\tau}\dot \phi dt & =& 2 \pi \int_0^{\tau} f(x) dt \label{eq:phi_int}\\
\int_0^{\tau}\dot x dt &=& - \alpha \int_0^{\tau} x(t) dt + k \int_0^{\tau} \sin(\omega t +\Delta \phi)x dt \label{eq:x_int}
\end{eqnarray}
where we introduced the phase shift $\Delta \phi$ between the entraining signal and the system (so that $\phi=0$ corresponds to the time of the pulse). This way, we can rewrite $\theta(t)=\omega t + \Delta \phi$.

Focusing on Eq. \ref{eq:x_int}, the left-hand side of this expression is $-r\alpha$ because of the periodicity of the system and the pulse of magnitude $r\alpha$ in $x$ at $t=0$. For the first term on the right-hand side, we use Eq. \ref{eq:f} to notice that $\int_0^{\tau} x dt = \int_0^{\tau} f(x) dt -\int_0^{\tau} \sigma(x) dt $, and from Eq. \ref{eq:phi_int} $\int_0^{\tau} f(x) dt =\frac{1}{2\pi}\int_0^{\tau}\dot \phi=1$. Given that $r=1$, the $r\alpha$ and $\alpha$ terms cancel out, and we get 
\begin{equation}
  \langle \sin(\omega t +\Delta \phi)x(t) \rangle=\frac{\alpha}{k} \langle \sigma(x(t)) \rangle \label{eq:deltaphipulse}
\end{equation}
which is reminiscent of Eq. \ref{eq:sigma}. Notice that in the limit of small $\alpha$ the right-hand side of Eq. \ref{eq:deltaphipulse} is expected to be rather small, which is consistent with the fact that we expect entrainment to be difficult in this limit.

To go further, one can integrate $x(t)$ to get (assuming $t=0$ corresponds to the pulse):
\begin{equation}
x(t)=\frac{\alpha e^{-\alpha t}}{1-e^{-\alpha \tau}}\exp{\frac{k}{\omega}[\cos (\Delta \phi)- \cos (\Delta \phi+\omega t)]}. \label{eq:ana_x_freq}
\end{equation}
Notice that $x$ depends on $\Delta\phi$ because the pulse happens at $t=0$. Substituting the value of $x$ into Eq. \ref{eq:deltaphipulse} provides an analytic expression allowing to define $\Delta \phi$ implicitly (see Appendix \ref{app:small_alpha_delta_phi} for the full expression). Like before, the stationary phase difference $\Delta \phi$ thus is a pure function of $\sigma$. The expression is highly non-linear and can not be inverted analytically, however, it can be solved numerically and compared to the steady-state value of the phase shift from explicit entrainment simulations, with perfect agreement, see Appendix \ref{app:small_alpha_delta_phi}, Fig. \ref{fig:figS8}.

To get a better sense of what happens, further approximations are needed. For low $\alpha$ Taylor expansion of the left-hand side of Eq. \ref{eq:deltaphipulse} gives
\begin{equation}\label{eq:lowalpha_sigma}
 \langle \sin(\omega t +\Delta \phi)x(t) \rangle = \frac{\alpha}{k} \left( I_0\left(\frac{k}{ \omega }\right) e^{\frac{k \cos (\Delta \phi )}{\omega }}-1 \right) 
\end{equation}
where $I_0(x)$ is the modified Bessel function of the first kind of order zero. This expression is proportional to $\frac{\alpha}{k}$ which thus simplifies with the right-hand side of Eq. \ref{eq:deltaphipulse}. One can use this Taylor expression to invert $\Delta \phi$, again with very good agreement, Appendix Fig. \ref{fig:figS8}.

We notice that $\Delta \phi$ depends only weakly on the detuning, indicative that the systems closely follow the entrainment signal in such a frequency coupling regime. This means that we can get a zero-order approximation of $\Delta \phi$ by taking its value for detuning $0$ (i.e. $\omega= 2\pi$) so that $ \langle \sigma \rangle =0$, which gives 
\begin{equation}\label{eq:delta_phi_freq_rough}
\Delta \phi \sim  \arccos\left(-\frac{2\pi}{k}\ln\left( \text{I}_0\left(\frac{k}{2\pi}\right)\right)\right)-\pi
\end{equation}

which is close to $-\pi/2$ for $k\sim 1$. For non-zero detuning, $\sigma$ is small because $x$ oscillates close to $1$, and $I_0\left(k/\omega\right)$ is order $1$ which is the reason $\Delta\phi$ does not change much with the detuning, consistent with the idea that the entrained system is poised to a precise phase difference in this regime, Appendix \ref{app:small_alpha_delta_phi} Fig. \ref{fig:figS8}.

\subsection{Large $\alpha$ regime}

So far we have implicitly assumed that $\alpha$ is small, which is necessary to have a weakly oscillating $x$, leading to a well-defined $f$ throughout the cycle. We are now relaxing this assumption to explore the limit of large $\alpha$.

In the averaging memory regime, taking in Eqs. \ref{eq:avg_syst} and \ref{eq:core} the quasi-static approximation $x\sim \frac{\dot \phi}{2\pi}$ gives $x \sim f(x)+\frac{k}{2\pi} H(\Delta \phi)$, which is the same equation as for the small $\alpha$ regime. So we do not expect any difference for the phase difference $\Delta \phi$ at entrainment for small and large $\alpha$, as indeed seen numerically in the first column of Fig. \ref{fig:fig9} B.

\begin{figure}
\includegraphics[width=8.5cm]{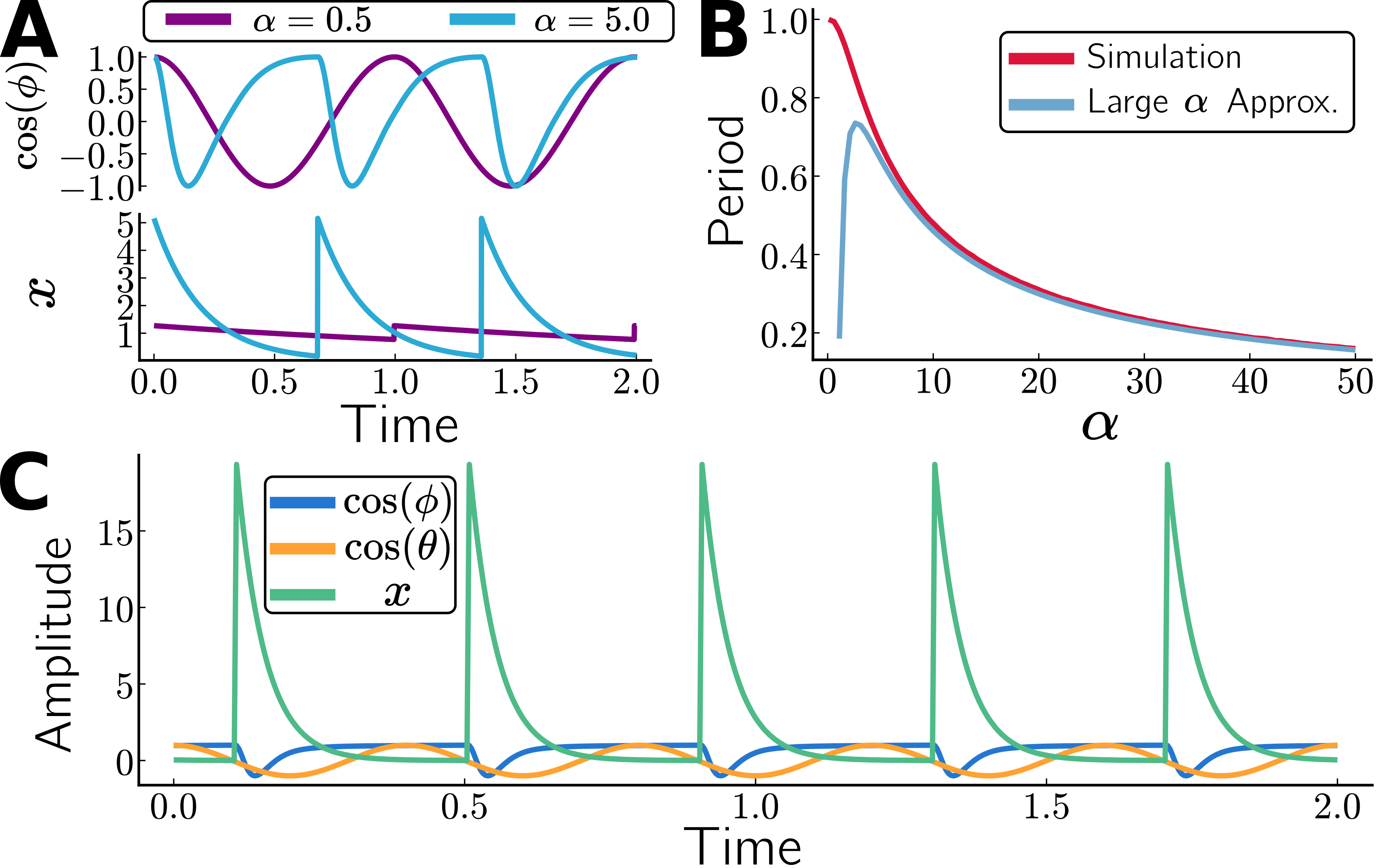}% Here is how to import EPS art
\caption{\label{fig:fig10}A) Unentrained time evolution of $\cos(\phi)$ and $x$ for the pulsatile systems, for different values of $\alpha$. Higher $\alpha$ leads to shorter stable unentrained periods (higher frequency) and asymmetric oscillations. B) Simulation and analytical approximation for the stable unentrained period for the Pulsatile system. Since this is for the unentrained system, this applies to both the Pulsatile memory-Core Phase entrainment and the Pulsatile memory-Frequency entrainment system. C) Example of entrainment of a high $\alpha$ system for Pulsatile memory-Core Phase entrainment system. The range of 1:1 entrainment gets shifted to shorter periods. $k=1$, $\alpha = 20$, $\omega/(2\pi)=2.5$.}
\end{figure}

However, for the pulsatile memory, the high $\alpha$ regime is different from the low $\alpha$ regime. The reason is that the magnitude of the pulse is proportional to $\alpha$. Because of this, $x$ becomes initially very big, but since the existence of the limit cycle requires that the average value of $x$ is $1$ (in the absence of entrainment), $x$ also needs to decay below $1$ significantly. Consequently, $f(x)$ varies much over one cycle. Thus, the system oscillates in a qualitatively different way: $\phi$ progresses rapidly right after the pulse but much more slowly when $x$ is decaying below $1$, as seen in Fig. \ref{fig:fig10} A (compare $\alpha=0.5$ where $\cos(\phi)$ is very sinusoidal to $\alpha=5.0$ where it looks more like a sawtooth).

In particular, the stable unentrained period depends on $\alpha$ in this regime, and can further be analytically computed. We assume that $x$ varies as $x(t) \sim \alpha e^{-\alpha t}$ for large $\alpha$. Assuming quick transitions between the plateaus of the sigmoid $\sigma$ we obtain the relation $\tau^* = 2 \frac{\log \alpha}{\alpha}$ (see Appendix \ref{app:deriv_stable_period} for derivation). Fig \ref{fig:fig10} B shows the comparison between this approximation and the measured period in simulations, with an excellent agreement for high enough $\alpha$.

Because of the non-uniform progression of $\phi$, entrainment properties are largely modified, and analytical results are more difficult to obtain. For core phase entrainment, for long entrainment periods, the entrainment phase does not depend much on detuning and is approximately $0$, Fig. \ref{fig:fig9}B, bottom middle. This is not surprising since for high $\alpha$, $\phi$ spends much time right before the pulse time $\phi=0$ as can be seen in Fig \ref{fig:fig10} A, and thus $\theta$ aligns with this point. For frequency coupling, there is now much more information on the global phase of the oscillator in the variable memory $x$ in this case, and remarkably, one recovers a plateauing dependency similar to the core phase entrainment at low $\alpha$, Fig. \ref{fig:fig9}B, bottom right. Notice however that, since $\dot \phi$ can no longer be considered uniform over a cycle for big $\alpha$, $\phi$ can no longer be directly taken as the "true" oscillator phase \cite{Kuramoto}. Such non-uniform phase increase is rather reminiscent of oscillators tuned close to an infinite period bifurcation \cite{francois2024} or the ERICA model for somitogenesis proposed in \cite{Sanchez2022}, with strong influences on the shape of the Phase Response Curve.

\section{Discussion}

In this work, we propose and study simple models of "unclocklike" oscillators with an internal memory variable. We are motivated by unexpected properties of the embryonic segmentation clock in response to entrainment \cite{Sanchez2022}, and by the more general idea that some biochemical memory variable $x$ is controlling the internal frequency $f(x)$ of the oscillator, while itself oscillating. We observe new entrainment properties that can not be captured with classical entrainment theory because it lacks such memory.

Formalizing the problem of frequency adaptation allows us to explore necessary conditions and to suggest new effects. We study two regimes: one with averaging memory and the other with pulsatile memory, and show they exhibit similar properties (in Supplement, we further show that multiple properties generalize as well to "soft" Gaussian pulsatile memories, Fig. \ref{fig:figS9}). Simple fixed point analysis shows that the frequency $f(x)$ of the phase oscillator $\phi$ has to be sublinear in the memory variable $x$, close to its natural frequency. It makes intuitive sense: for instance, if the frequency of the system were to be slightly increased externally by some transient perturbations, it then has to go down closer to its natural frequency, thus ensuring homeostasis. 

The effect of such frequency adaptation is mostly visible in entrainment experiments. It first allows for a considerably bigger entrainment range with broad Arnold tongues, Fig. \ref{fig:fig4}. Furthermore, in all models considered, the entrainment phase is always a function of the non-linearity of the frequency, captured by the function $\sigma(x)=f(x)-x$. This observation is of practical importance. Concretely, an experimentalist could vary the frequency of entrainment $\omega$, measuring the entrainment phase difference $\Delta \phi$ to infer properties of both the coupling functions $H$ and non-linearity in the frequency $\sigma$, which quantifies the feedback between the memory variable and the frequency of the oscillator. Case in point, in the vertebrate segmentation clock \cite{Sanchez2022}, the shape of $H$ was first inferred, then the plateauing of $\Delta\phi$ suggested both the existence of such feedback and its functional form, i.e. $f(x)=x\pm c_1$ for high and low detuning. Systematic explorations of different biological oscillators might reveal different types of $\sigma$ corresponding to different internal controls.

We also demonstrate two new hysteretic effects associated with such models. Exploring the full entrainment range is only possible with a smooth change of frequencies, allowing the memory variable $x$ to change adiabatically. Conversely, if one suddenly changes the frequency of entrainment close to the boundary of the entrainment range, $x$ can not adapt rapidly, so entrainment is not always possible. This is the first new hysteresis type: the existence of entrainment with high detuning depends on the past history of $x$, Figs. \ref{fig:fig5}-\ref{fig:fig6}. Another hysteretic effect is bistability in the \textit{intrinsic} frequency of the oscillator, depending on the value of $x$ which can be captured by bi-dimensional maps, Fig. \ref{fig:fig7}. In biological contexts such as the segmentation clock, one could imagine observing the two hysteretic effects within one single experiment by carefully playing with the timing of entrainment. For instance one could start from a system entrained at the intrinsic frequency, then slowly change the entrainment frequency while staying in the 1:1 tongue, ensuring a big change of $x$ to adapt to entrainment. Then, one could pause the entrainment signal, to give enough time for  $x$ to revert the system to its intrinsic frequency, then start entrainment again at the same frequency before the pause, to reach 1:2 entrainment. 

Entrainment appears more difficult through $x$ than through $\phi$, possibly because phase effects on $x$ are averaged by $f$, leading to a situation closer to overdrive suppression where the behavior is best explained by a simple change of one internal parameter \cite{Zheng2015}. That said, if some processes or variables directly control the period of a biological oscillator, they are likely to themselves oscillate, so a (weak) phase dependency is not unexpected, and would thus allow for entrainment. This raises the question of dual entrainment, where two signals act on different parts of the biochemical network, a situation with direct relevance for the segmentation clock where there are clear metabolic inputs \cite{diaz2023metabolic, miyazawa2024} but, maybe more importantly, with clinical relevance in chronobiology, e.g. the case of treatment modulated by the circadian clock \cite{kim2013modeling, kim2019systems}.

%In the segmentation clock entrainment context, one could imagine combining the two hysteretic effects within one single experiment: one could slowly increase (or decrease) decrease the entrainment frequency by a factor of $2$, ensuring the intrinsic period is double, before suddenly reverting to the intrinsic frequency of the oscillator, thus ensuring 1:2 entrainment. 

In the segmentation clock context, entrainment experiments have been done via perturbations of Wnt, Notch \cite{Sonnen2018} and metabolism acting via Wnt \cite{miyazawa2024}. The comparison of entrainment ranges and entrainment phases for each of those situations might help to uncover qualitative differences in entrainment properties, possibly identifying pathways responsible for different organization levels in the biological "unclock" considered (i.e. core clock phase $\phi$ vs. memory variable $x$). It is already known that Wnt and Notch relative phases have a functional role in differentiation \cite{Sonnen2018}. Thus, a single phase cannot fully describe the segmentation clock, which is "unclocklike". It has further been suggested that a second oscillator (possibly Wnt) changes the frequency of the Notch oscillator \cite{Lauschke2013}. Interestingly, Wnt oscillations are experimentally much more synchronized within the embryo than Notch \cite{Sonnen2018}, a possible indication that Wnt could indeed implement and encode some averaging of Notch oscillators, thus possibly playing a role similar to the memory variable $x$. Adjustments of memory $x$ could play a role in the versatility of the segmentation clock to multiple perturbations (e.g. changes of temperature \cite{zhang2022fgf8} or metabolism \cite{diaz2023metabolic,miyazawa2024}, or scaling \cite{seleit2024modular}).

Progress in quantitative biology and real-time imaging has opened the door for systematic exploration, modeling, and understanding of biological dynamics. In some cases, one can directly relate biology to well-known, classical mathematical descriptions (e.g. landscapes \cite{Rand2021}), and theoretical modeling involves mapping mathematical control parameters to biological handles. However, dynamical systems in biology are often more versatile and adaptive than classical physics ones. A complete understanding requires specific hypotheses tied to uniquely biological properties, such as homeostasis, adaptation, or more generally computation \cite{koseska2017}. Memory variables like those proposed here explain how the timescale of biological oscillators can adjust to external constraints, possibly contributing to the labile tempo of embryonic development \cite{EBISUYA2024,manser2023}. Well-defined perturbations, such as entrainment, reveal properties of the internal biological feedbacks like memory variables and can be tested on any biological oscillators. This opens the door for the building and the exploration of new classes of models like we do here, and further interactions between theory and experiments.

%Motivation: entrainment of somite clock with unusual features + two oscillator framework. Deserves new models.

%We propose frequency adaptation, via a memory variable.

%we recapitulate puzzling features of the entrainment experiments: very broad Arnold tongues but also phase difference information. 

%Experiments to do :
%hysteresis of frequency: change smoothly entrainment period and see what happens, vs direct entrainment
%bistability: go to a very low frequency, then revert to intrinsic frequency: do we see 2:1? Is there multistability of frequencies? This would be the smoking gun for feedbacks 

%entrainment of two oscillators: smooth phase or conflicts? Something else?

%On a theory side :

 %So in a more realistic/complex model, one could explicitly model coupled differential equations corresponding to a biochemical oscillator in lieu of $\phi$ (e.g. a delayed oscillator \cite{Lewis2003}); our assumption here is that one variable in the biochemical network ($x$)  influences the frequency but also is controlled by the core oscillator (encoded by $\phi$). 

%what could be x ? 
%average/reference oscillator

%could explain global frequency adaptation and other features such as scaling.

%From a broader standpoint, such versatility of entrainment behaviours is due to 

%We expect biology to be much more adaptive, to have ranges of homeostasis, which motivates new models with new properties as described here.

\begin{acknowledgments}
We thank Alexander Aulehla, Hidenobu Miyazawa, Jona Rada, and members of the Aulehla and Francois's groups for multiple discussions.
\end{acknowledgments}

% s\nocite{*}

\bibliography{Biblio}% Produces the bibliography via BibTeX.

\newpage
\appendix

\renewcommand\thefigure{S\arabic{figure}}
\setcounter{figure}{0}  

\section{Brief summary of classical entrainment theory} \label{app:classical_entrainment}

In this section, we briefly summarize aspects of classical entrainment theory of limit cycles to derive equations used for comparison in the main text.
The phase of a clock is a zero mode on a limit cycle \cite{pikovsky2001}, and it can be shown that at the lowest order, assuming weak coupling, the dynamics of the entrained oscillator (phase $\phi$, intrinsic frequency $\omega_i$) are influenced by the entraining periodic signal (phase $\theta$, frequency $\omega$ ) in the following way :
\begin{equation}
\frac{d\phi }{dt}=\omega_i+H(\theta-\phi) \label{eq:Hdefinition}
\end{equation}
where $H$ is a function of phase (difference) that depends only on the shape of the flow of the entrained oscillator close to the limit cycle. For the segmentation clock, although DAPT pulses are only weakly influencing the oscillation \footnote{in the sense that it requires multiple cycles for the clock to be entrained and that each pulse leads to only moderate phase change}, $H$ is strongly shifted depending on external oscillator frequency, which is not expected from classical theory since $H$ does not depend on $\omega$ in Eq. \ref{eq:Hdefinition}. Practically, this leads to "Arnold tongues" for entrainment much larger than usual \cite{Sanchez2022}. Another inconsistency with the classical entrainment theory is the unusual behavior of the entrainment phase as a function of detuning \cite{Granada2013}. Assuming that the external oscillator $\theta$ has frequency $\omega$, from Eq. \ref{eq:Hdefinition} entrainment occurs if 

\begin{equation}
\omega_i-\omega+H(\theta-\phi)=0 \label{eq:freq_entrained}
\end{equation}
This means the range of entrainment is defined by maxima and minima of $H$, where the derivative $H'$ is $0$.
%which imposes a condition on the detuning $\omega-\omega_0$ as a function of $H$. For instance, to have entrainment, one necessarily needs $\left |\frac{\omega-\omega_0}{c}\right|<Max(H)$, and then 
Furthermore, the relative entrainment phase of the oscillator $\phi_e$ with respect to the entrainment signal $\theta$ is thus of the form 
\begin{equation}
\phi_e-\theta=H^{-1}\left(\frac{\omega_i-\omega}{c}\right) \label{eq:phase_entrained}
\end{equation}
then, $d\phi_e/d\omega$ is proportional to $1/H'$ at the range of entrainment, so should become infinite (as illustrated in Fig. \ref{fig:fig8}). Surprisingly, the entrainment phase for the segmentation oscillator plateaus for very small and very large detuning, thus completely inconsistent with the classical entrainment theory.

\section{Numerical Integration}\label{app:numerical_integration}

\begin{figure}
\includegraphics[width=8.5cm]{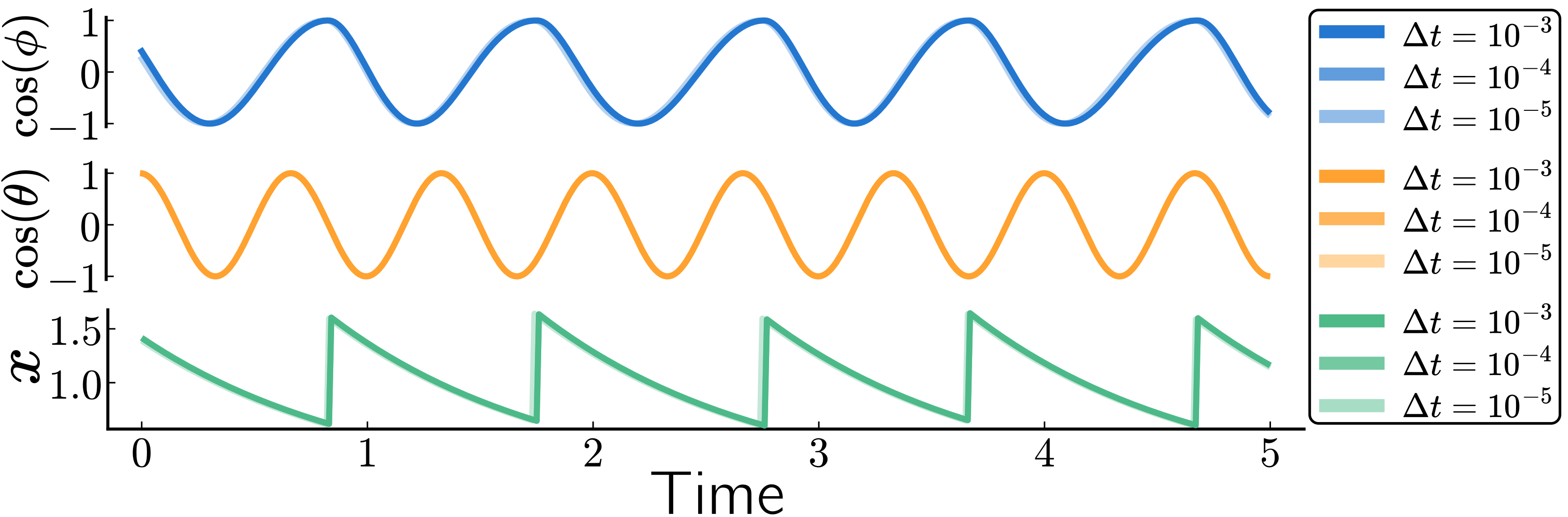}
\caption{\label{fig:figS1}Testing the numerical stability of the pulsatile memory-core phase entrainment system by varying the time step of the integration. A transient time of 500 time units was used before the displayed time range. We see minimal disparity between different time steps. $\omega = 1.5/(2\pi)$. $k=0.5$. $\alpha = 1.0$.}
\end{figure}

All the simulations presented in this paper were done in the Julia programming language \cite{julia}, using the \texttt{DifferentialEquations.jl} package \cite{rackauckas2017differentialequations}.

The notebook and other scripts to generate the data for all the figures in this paper (except conceptual Figs. \ref{fig:fig1} and \ref{fig:figS10}) is available at the following repository: \href{https://github.com/cmdenis/2024-frequency-memory}{https://github.com/cmdenis/2024-frequency-memory}

We used the built-in \texttt{Euler} solver from \texttt{DifferentialEquations.jl}. To properly integrate Dirac functions, we define a "callback" that adds the value $r\alpha$ to the $x$ variable whenever the solver detects that $\phi$ modulo $2\pi$ is equal to 0. The stability of the solutions was checked by varying the time step for integration in Fig. \ref{fig:figS1} as well as through comparison with some analytical results presented in this paper. Notice that Julia plotting tools connect lines between discontinuous data points, but we checked that our time-course with Dirac pulses is truly discontinuous.

\section{Varying $r$}
\label{app:var_r}
\begin{figure}
\includegraphics[width=8cm]{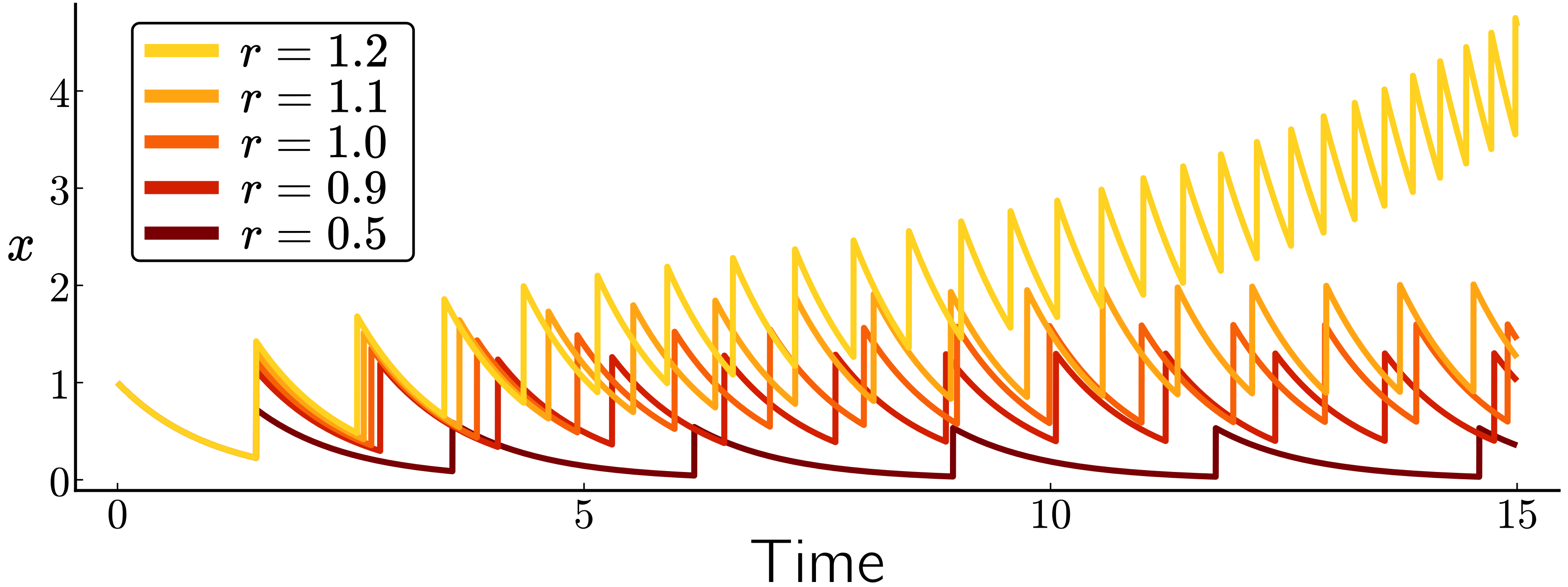}% Here is how to import EPS art
\caption{\label{fig:figS2} Effect of different ratios $r$, taking $\alpha=1$. Note that the system blows up (and speeds up) if $r$ is too big.}
\end{figure}

Varying the size of the pulse, through the $r$ parameter in the pulsatile memory system can give stable unentrained pulsation. However, it was seen numerically that high values of $r$ can start diverging numerically. For this reason, as well as mathematical convenience, we focus our attention on systems with $r=1$. Unentrained scenarios for $r\neq 1$ are presented in Fig. \ref{fig:figS2}.

\section{Arnold Tongues for the Averaging memory-Core phase entrainment System}
\label{app:tongue-average}
\begin{figure}
\includegraphics[width=8cm]{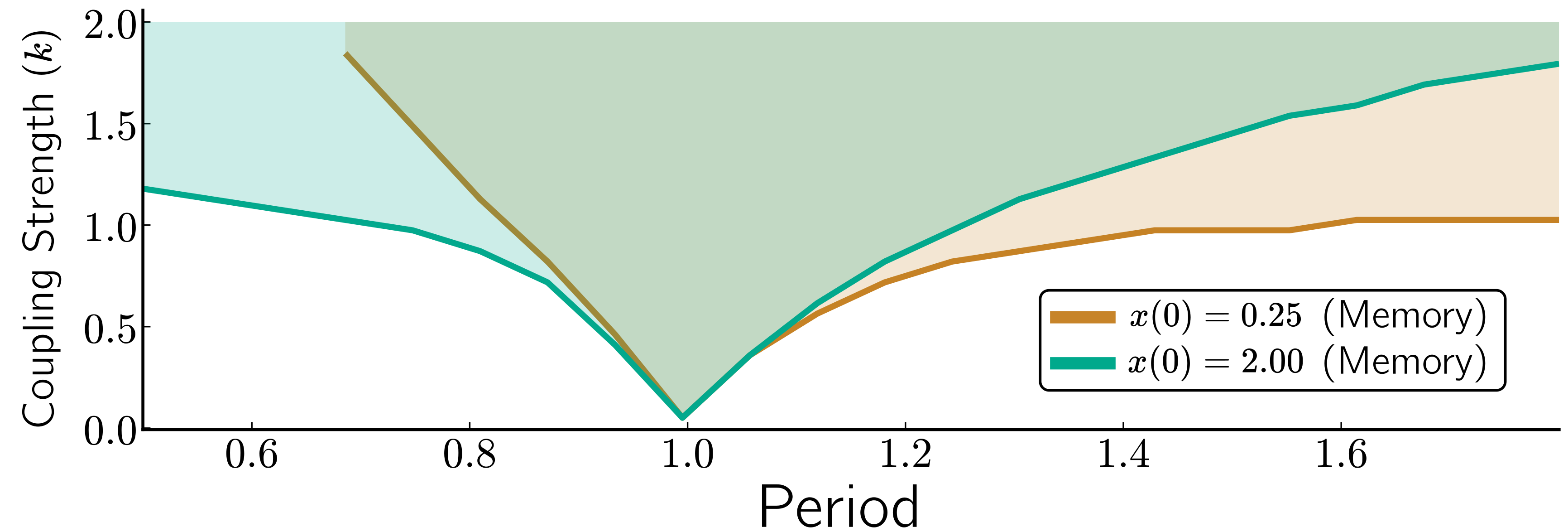}% Here is how to import EPS art
\caption{ 1:1 Arnold tongue for the Averaging-Core phase system. Different initial conditions for the $x$ variable yield tongues stretching to shorter or longer periods. $\alpha = 0.1$}\label{fig:figS3}
\end{figure}

Just like the Pulsatile-Core phase system, we can obtain Arnold Tongues for the Averaging-Core phase system. We see the same behavior as in the pulsatile system when using different initial conditions for the $x$ variable in Fig. \ref{fig:figS3}.

\section{Hysteresis for the core phase coupling system}
\label{app:hysteresis}
\begin{figure}
\includegraphics[width=8.5cm]{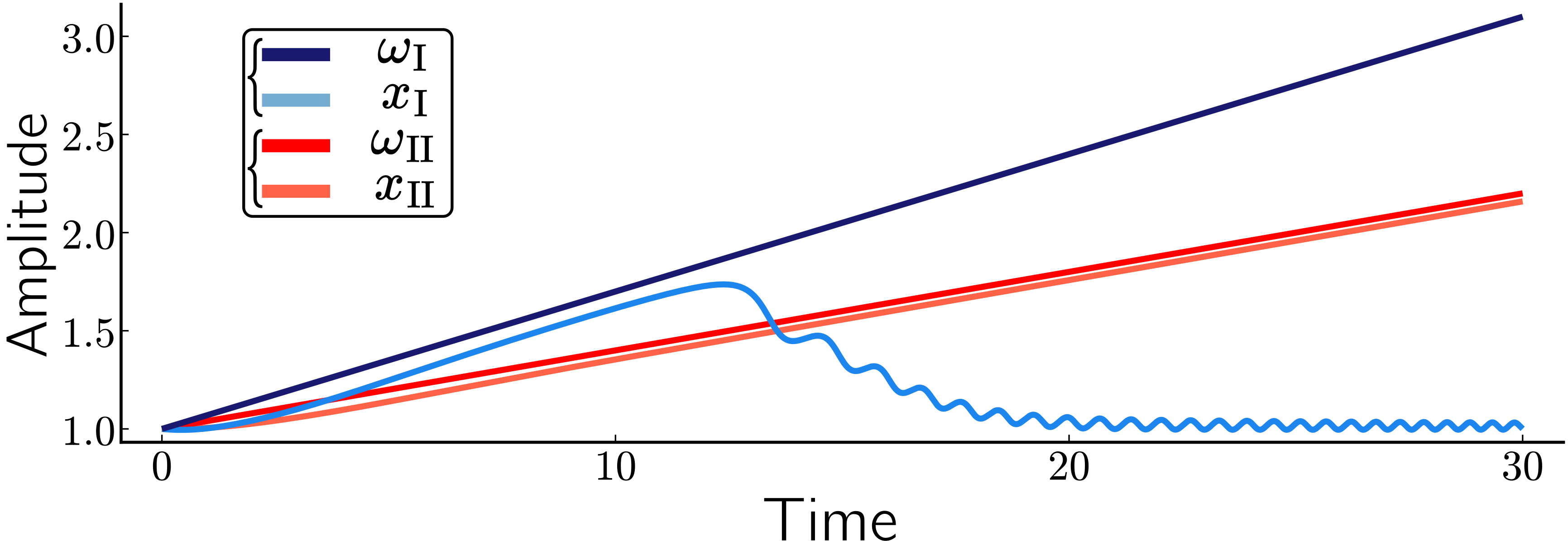}
\caption{\label{fig:figS4} Example runs for the Averaging-Core phase system being entrained and unentrained after ramping the entraining frequencies. For an entrainment strength of $k=1.50$. Only the entraining frequencies ($\omega_\text{I}$ and $\omega_{\text{II}}$) and the internal frequency ($x_\text{I}$ and $x_{\text{II}}$) are shown. $k=1.5$, $\alpha = 1$.}
\end{figure}

Hysteresis is noticed in the core phase coupling system. When varying the frequency of the entraining oscillator slow enough, we can entrain in a very wide range of frequencies. In fact, if the frequency is varied slowly enough and the amplitude of coupling is strong enough, no bound on the maximum frequency of entrainment was observed. Fig. \ref{fig:figS4} shows some quantitative results regarding the ramping of the frequency of the system.

\section{Comparison of the maps with the classical system}

We can create the same maps presented in figure \ref{fig:fig7} for the classical system by simply setting the coupling function $f$ to $1$ in the pulsatile memory system. This way we still have pulses and can draw the map at each pulse, but preserve the same dynamics as for the classical model. We compare those maps for the classical model with the ones for the pulsatile-core phase system in Figure \ref{fig:figS7}.

\begin{figure}
\includegraphics[width=8.5cm]{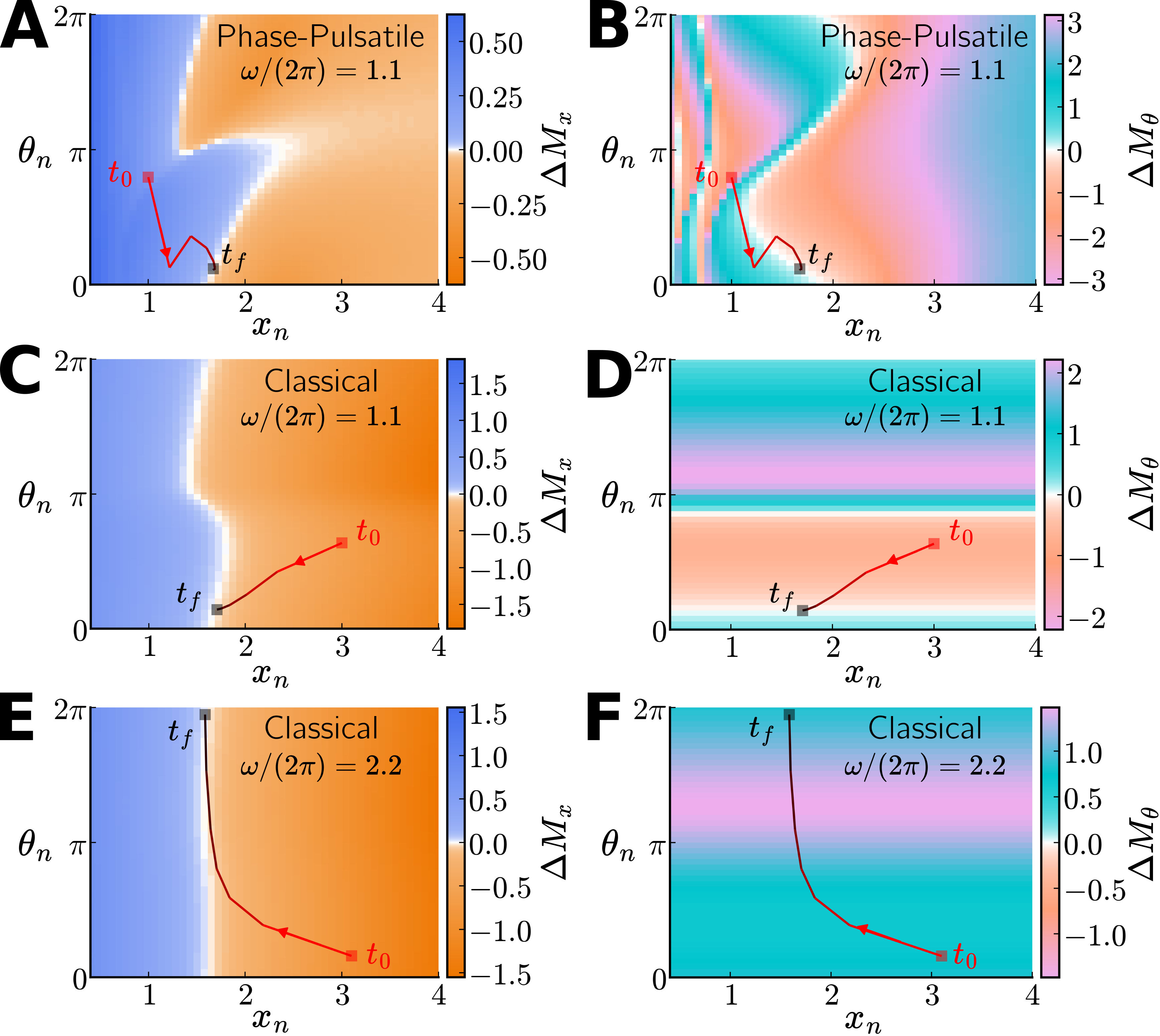}% Here is how to import EPS art
\caption{Comparison of the maps for $x_n$ and $\theta_n$, between the classical system and the pulsatile memory-core phase entrainment system. $k=1.5$. $\alpha = 1.0$. A-B) Maps for the pulsatile memory-core phase entrainment system with $\omega/(2\pi)=1.1$. C-D) Maps for a classical system without memory retrieval $\omega/(2\pi)=1.1$. $k=1.5$. E-F) Maps for a classical system without memory retrieval $\omega/(2\pi)=2.2$. This map does not feature a stable fixed point, unlike the other subplots. The trajectories keep drifting in phase.}\label{fig:figS5}
\end{figure}

\section{Variation of asymptotic behavior of the coupling function}
\label{app:asymptotic}
\begin{figure}
\includegraphics[width=8cm]{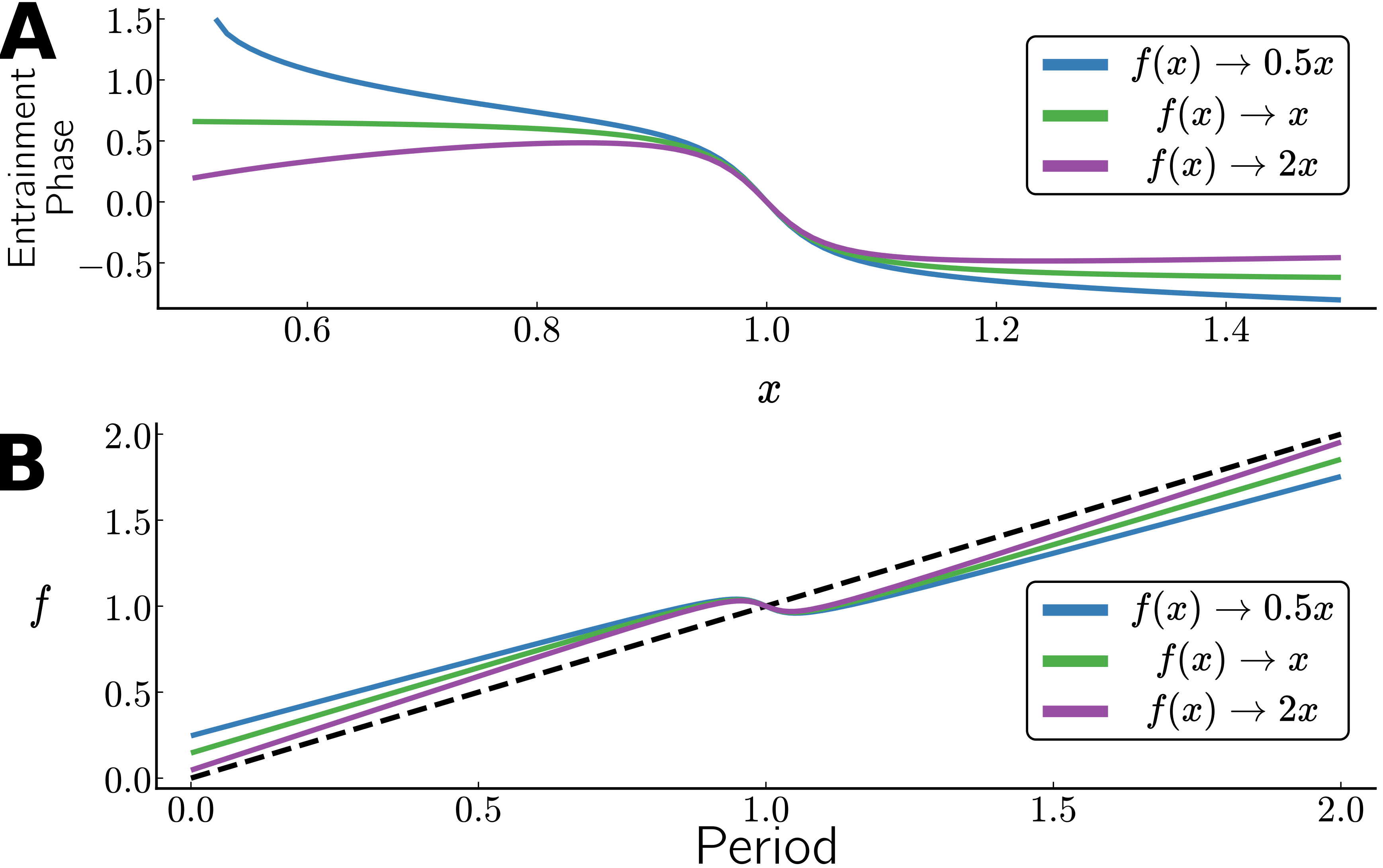}% Here is how to import EPS art
\caption{\label{fig:figS6} A) Coupling functions with different asymptotic behavior. For high or low $x$, curves go like $0.5x$, $1x$ and $2x$. $k=1.5$. B) Analytical curves using equation \ref{eq:entrain_phase_avg} for the entrainment phase curves resulting from the frequency curves $f$ presented in panel A, using core phase coupling with $H=\sin$.}
\end{figure}

We assumed that $f\sim x$ for $x$ away from 1, consistent with experiments. Such a precise constraint might however not always be realistic. However, we can relax this assumption to show that the system still possesses a characteristic flattening of its entrainment phase curve, as illustrated with numerical simulations in Fig. \ref{fig:figS6}. 
If $f\sim ax$, then for $a>1$, there will be other points where $f(x)=x$ for $x\neq 1$, this means that there will be an infinite derivative at the end of the phase curve, but if $a$ is close to one, it will still be preceded by a flattening. 
If $a<1$, then the phase curve will not only flatten but also change the sign of its derivative past a maximum or minimum.

\section{Entrainment Phase for the Pulsatile Memory-Core phase entrainment system}
\label{app:pulse_entrainment}
We consider the pulsatile regime

\begin{equation}
    \begin{cases}
\dot{\phi}=2 \pi f(x)+k H \left(\theta - \phi\right)\\
\dot{x} = -\alpha x + r\alpha \delta\left(\textbf{mod}(\phi,\ 2\pi)\right).
\end{cases}
\end{equation}

Integrating the $\phi$ variable over one cycle

\begin{equation}
    \frac{1}{\tau}\int_0^\tau \dot{\phi}\ dt =\frac{1}{\tau}\int_0^\tau \left(2 \pi f(x)+k H (\theta - \phi) \right)\ dt.
\end{equation}

We know that, $\frac{1}{\tau}\int_0^\tau \dot{\phi}\ dt = \frac{2\pi}{\tau}=\omega$. Hence, we get 

\begin{equation}
  \omega=2\pi\langle f(x)\rangle  + k\langle H(\Delta \phi)\rangle.
\end{equation}
where averages are computed over one cycle.

The "Dirac" pulse in $x$ only occurs once every cycle, at $\phi=2\pi$. This means that right after this pulse, the equation for $x$ is linear $\dot x=-\alpha x$ and can thus be fully integrated until the next pulse :

\begin{equation}
    x(t)=e^{-\alpha  t } x_0 \label{eq:x_A}
\end{equation}

where $t$ is time, $\alpha$ is the decay rate and $x_0$ is the initial value of $x$ right after the pulse is applied, corresponding to the beginning of the cycle $t=0$.
If the system is entrained, the behavior of $x$ is periodic, of period $\tau$, which is the same as the entraining signal. Thus $x_0=x(\tau)+r\alpha=x_0 e^{-\alpha \tau}+r\alpha$, which expresses the fact that there is a pulse of magnitude $r\alpha$ at $\tau$ to bring $x$ up to the initial value after the pulse $x_0$. From this, we get:
\begin{equation}
x_0 = \frac{r\alpha}{1-e^{\alpha  \tau }}. \label{eq:x_0}
\end{equation}

which allows us to fully express $x$ as a function of time within one cycle (assuming $t=0$ corresponds to the time of the pulse) :

\begin{equation}
    x(t) = r\alpha\frac{e^{-\alpha t}}{1-e^{-\alpha \tau}}.
\end{equation}

We then define $\chi$ as

\begin{equation}\label{eq:chi}
    \chi = \omega \int_0^\tau f\left(r\alpha\frac{e^{-\alpha t}}{1-e^{-\alpha \tau}}\right)\ dt.
\end{equation}

which can not be computed analytically without further assumptions.
Assuming now that the phase difference $\Delta \phi$ between the two oscillators is roughly constant for the duration of the cycle (an assumption that we relax below), using the expression for $\phi$ in Eq. \ref{eq:pulse_syst}, we get

\begin{equation}
    \Delta \phi = H^{-1}\left(\frac{ \omega -  \chi}{k} \right). 
\end{equation}

We can also use this expression to estimate the stable entrained frequency, $\omega^*$ by looking at the case where $\Delta \phi = 0$ (assuming $H(0)=0$ corresponds to the stable coupling. This yields the relation

\begin{equation}
    \omega^* = \omega\int_0^\tau  f\left(r\alpha\frac{e^{-\alpha t}}{1-e^{-\alpha \frac{2\pi}{\omega^*}}}\right) dt.
\end{equation}

Notice that in the limit of small $\alpha$ $\frac{e^{-\alpha t}}{1-e^{-\alpha \tau}}\rightarrow  \frac{r\alpha}{\alpha\tau^*}$, so that 
\begin{equation}
\omega^* = \omega f(\frac{r}{\tau^*}) 
\end{equation}
For $r=1$ and $f(1)=1$, the intrinsic period $\tau^*=1$ is solution as desired.

One can also study what happens for high and low detuning.
Assuming $f(x)=x+\sigma(x)$ where $\sigma$ is a sigmoidal function, and assuming saturation far from the intrinsic entrainment frequency, we can make the approximation 

\begin{equation}\label{eq:f_approx}
    f(x) \approx x + C.
\end{equation}

Hence, looking back at the expression for $\chi$ in the equation \ref{eq:chi} and using the approximation in \ref{eq:f_approx} we get

\begin{equation}
    \chi =  \omega\int_0^\tau \left( r\alpha\frac{e^{-\alpha t}}{1-e^{-\alpha \tau}} + C\right) dt
\end{equation}

which yields

\begin{equation}
    \chi =  \frac{r\alpha}{\alpha}\omega + 2\pi C.
\end{equation}

so that 
\begin{equation}
\frac{d \chi}{d \omega} \sim \frac{r\alpha}{\alpha}=r \label{eq:asymp_chi}
\end{equation}

as claimed in the main text.

\section{Varying the entrainment coupling function for the pulsatile memory-frequency entrainment system}\label{app:var_coup_freq}

\begin{figure}
\includegraphics[width=8cm]{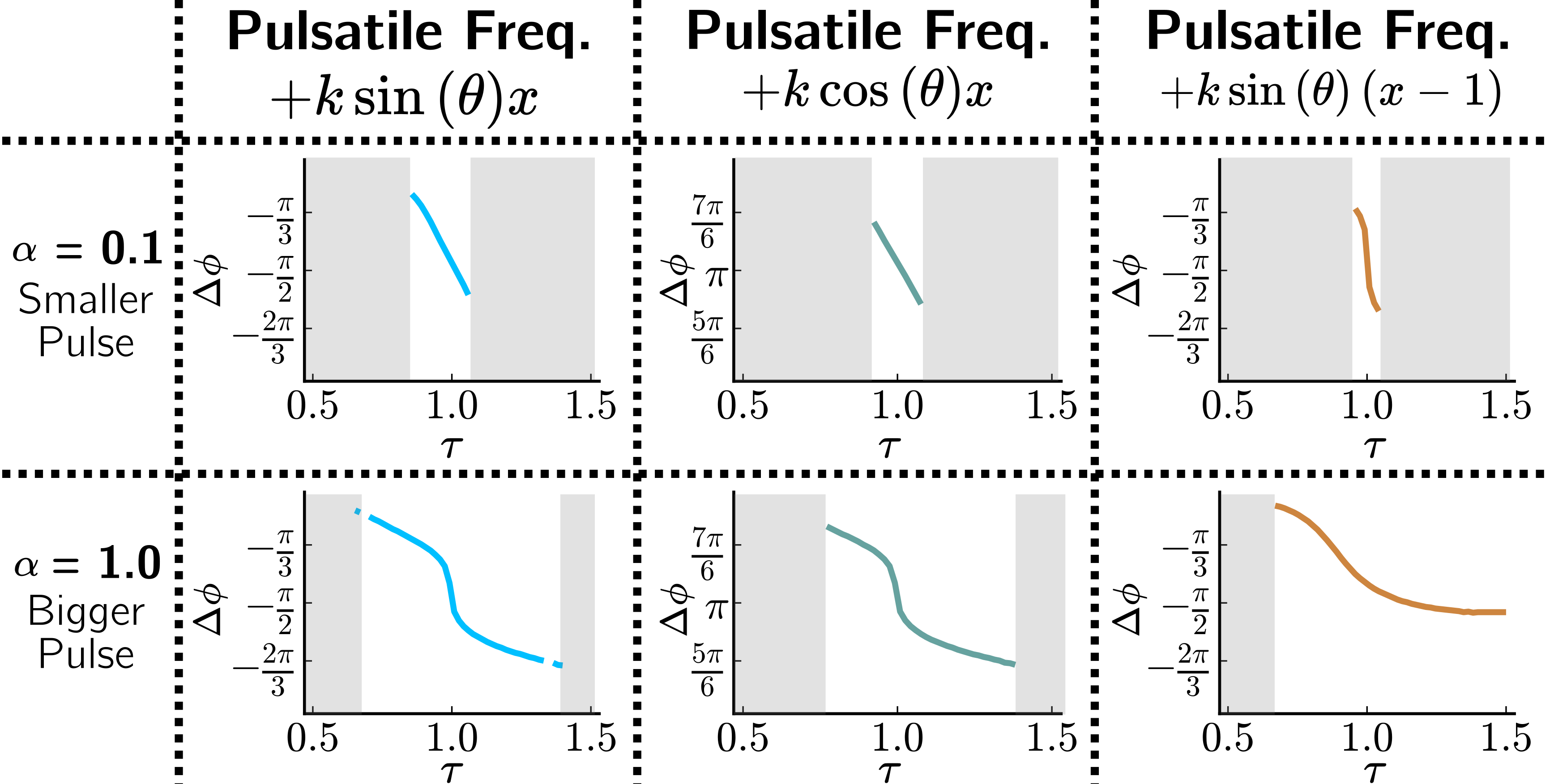}% Here is how to import EPS art
\caption{\label{fig:figS7} Phases of entrainment for different coupling terms in the Pulsatile memory-Frequency entrainment system with $k=1.5$.}
\end{figure}

We can create variations of the Pulsatile memory-Frequency entrainment model by modifying the coupling function used for entrainment between $\theta$ and $x$. In the main text, we presented a coupling function of the form $+k\sin \left( \theta \right) x$. Other variations such as $+k\cos \left( \theta \right) x$ and $+k\sin \left( \theta \right) (x-1)$ also yield entrainment. The phase of entrainment for these alternative systems is compared in Fig. \ref{fig:figS7}, with qualitative behaviors similar to the case presented in the main text.

\section{Phase of entrainment for large $\alpha$ in pulsatile memory-frequency entrainment system}
\label{app:small_alpha_delta_phi}

\begin{figure}
\includegraphics[width=8.5cm]{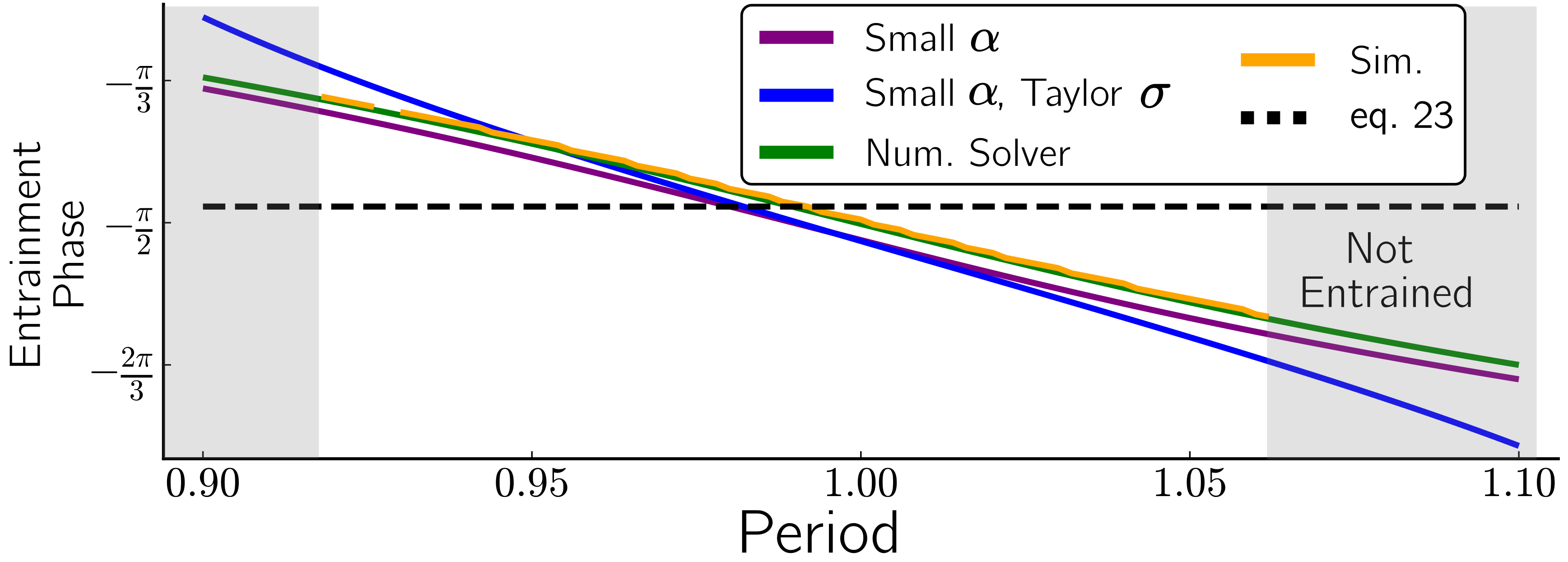}% Here is how to import EPS art
\caption{\label{fig:figS8} Phase of entrainment for the Pulsatile memory-Frequency entrainment system using different approximations. The yellow curve corresponds to the actual value of $\Delta \phi$ measured from full integration of the differential equations. The green curve corresponds to the numerical solving of the implicit Eq. \ref{eq:explicit_sigma_eq} for $\Delta \phi$, which agrees perfectly with the simulations. The purple curve corresponds to the numerical solving of Eq. \ref{eq:explicit_sigma_eq} after approximating the trajectory of $x(t)$ with a Taylor expansion for small $\alpha$. The blue curve corresponds to the numerical solving of Eq. \ref{eq:explicit_sigma_eq} while linearizing $\sigma$ around 1 and using a Taylor expansion in $\alpha$ for $x(t)$. The dotted black curve corresponds to the constant value Eq. \ref{eq:delta_phi_freq_rough}. For those simulations, we used $k=0.9$, $\alpha = 0.1$, $c_1 = 8$, $c_2=4$.}
\end{figure}

We have shown analytical results for the phase of entrainment as a function of the period of the entraining oscillator. The phase of entrainment for the pulsatile memory-frequency entrainment system can be obtained in the limit of small $\alpha$. However, it was seen numerically that smaller values of $\alpha$ reduce the range of entrainment, as in Figs. \ref{fig:fig9} and \ref{fig:figS7}. We plot the results from section \ref{sec:entrain_memory} in Fig. \ref{fig:figS8}.

For the entrained pulsatile memory-frequency entrainment system, we can write exactly $x$ as a function of time using a similar reasoning as the one presented in section \ref{app:pulse_entrainment}. We take the origin of time $t=0$ right after the pulse: until the next pulse, the dynamics of $x$ obey the linear equation$\dot{x}(t)= -\alpha x + k sin(\theta) x$. Assuming the system is entrained, we right $\theta = \omega t +\Delta\phi$, and using the separation of variables, to integrate the linear equation for $x$, we obtain:
\begin{equation}
    x(t) = C \exp{ \left(-\frac{k}{\omega} \cos (\Delta \phi -t \omega )- \alpha  t \right)}
\end{equation}

To compute the constant $C$ we apply the periodic boundary condition for entertainment, taking into account the periodic pulse happening at time $t=\tau$, i.e. we have $x(0)=x(\tau) + \alpha$. This yields

\begin{equation}
    C = \frac{\alpha  \exp{\left( \frac{k}{\omega}\cos (\Delta \phi )\right)}}{1 - \exp{\left(-\alpha \tau \right)}}
\end{equation}

Hence, we have

\begin{equation}
    x(t) = \frac{\alpha  e^{\alpha t}}{1 - e^{-\alpha \tau }}   \exp{ \left( \frac{k}{\omega} \left( \cos (\Delta \phi ) -\cos (\Delta \phi -t \omega ) \right) \right)}
\end{equation}

Using this result and expanding Eq. \ref{eq:deltaphipulse}, we can obtain an exact analytical implicit solution for $\Delta \phi$ in the pulsatile memory-frequency entrainment system:

\begin{multline}\label{eq:explicit_sigma_eq}
    -\frac{\alpha}{k} \int_0^{\tau} \sigma \left(\frac{\alpha e^{-\alpha t}}{1-e^{-\alpha \tau}}\exp{\frac{k}{\omega}[\cos (\Delta \phi)- \cos (\Delta \phi+\omega t)]}\right)  dt \\ = \int_0^{\tau}\sin (\Delta \phi + \omega t ) \frac{\alpha e^{-\alpha t}}{1-e^{-\alpha \tau}}\exp{\frac{k}{\omega}[\cos (\Delta \phi)- \cos (\Delta \phi+\omega t)]} \, dt.
\end{multline}

In other words, given $\tau = 2\pi/\omega$, $\alpha$ and $k$, one can solve implicitly for $\Delta \phi$. Assuming small $\alpha$ gives

\begin{equation}\label{eq:pulse_freq_x_small_alpha}
    x_\alpha(t) =\left(\alpha  \left(\frac{1}{2}-\frac{t}{\tau }\right)+\frac{1}{\tau }\right) e^{\frac{k \cos (\Delta \phi )-k \cos (\Delta \phi +t \omega )}{\omega }}.
\end{equation}

with this, we can solve the right-hand side of Eq. \ref{eq:explicit_sigma_eq} exactly and obtain Eq.
\ref{eq:lowalpha_sigma}. Using Eq. \ref{eq:pulse_freq_x_small_alpha}, one may further simplify the left-hand side of Eq. \ref{eq:explicit_sigma_eq} by Taylor expanding $\sigma(x)\approx-\frac{c_2}{c_1}(x-1)$ near $x=1$. This result is compared to our simulation and the linear approximation mentioned in the main text in Fig. \ref{fig:figS8}, with excellent agreement.

\section{Analytical computation of stable period for large $\alpha$}\label{app:deriv_stable_period}

We want to find $\omega^*=1/\tau^*$ such that

\begin{equation}
    \omega^* =\langle f(x)\rangle = \langle \sigma(x)\rangle + \langle x \rangle =  \langle \sigma(x)\rangle + \omega^*
\end{equation} 

Which means we need $\langle \sigma(x)\rangle= 0$. Since $x$ varies quickly between the two saturation plateau of $\sigma$, we approximate $\sigma$ as a piecewise function:

$$
\sigma(x) =\begin{cases}
\sigma_+,\ \text{if $x\leq1$}\\
\sigma_-,\ \text{if $x>1$}\\

\end{cases}
$$
If we find $t^*$ where $x(t^*)=1$, then we could write

$$ \langle \sigma (x) \rangle = \frac{1}{\tau^* }\left(\sigma_-t^* +\sigma_+(\tau^* - t^*) \right)$$

We can find $t^* = \frac{\log \alpha}{\alpha}$ , which gives us

$$ \frac{1}{\omega^*}=\tau^*=\frac{\log \alpha}{\alpha}\frac{ (\sigma_--\sigma_+)}{\sigma_-} $$
And if we assume $\sigma$ to be symmetric, such that $\sigma_- = -\sigma_+$, then we get 

\begin{equation}
    \frac{1}{\omega^*}=\tau^*=2\frac{\log \alpha}{\alpha}.
\end{equation} 

This result is presented and compared to the simulations in the main text, Fig. \ref{fig:fig10}.

\section{Softer Pulse}

\begin{figure}
\includegraphics[width=8cm]{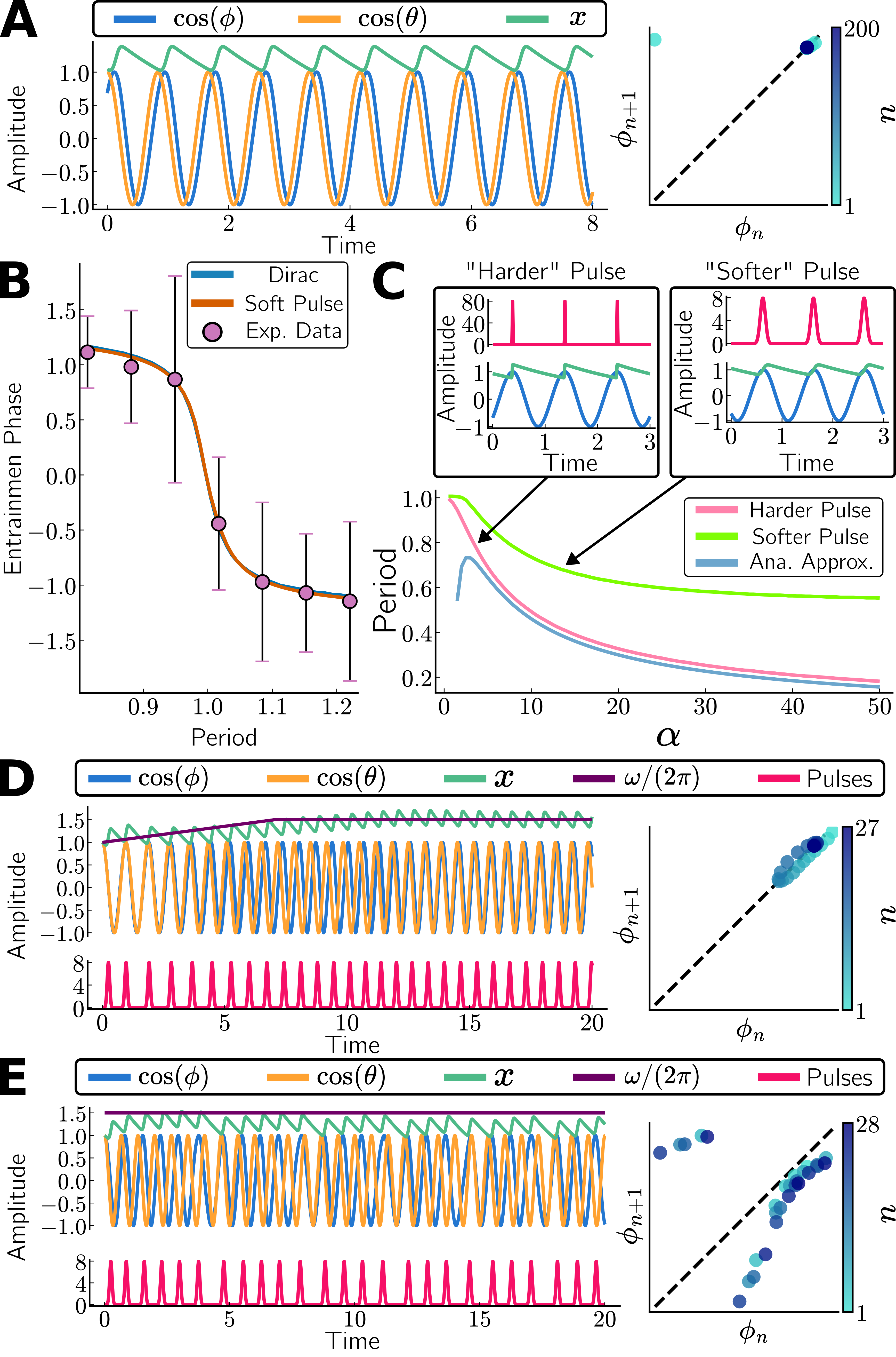}% Here is how to import EPS art
\caption{A) Example of entrainment using a Gaussian pulse instead of a Dirac pulse. Left: Trajectories of the entrained system. Right: Stroboscopic map of the trajectories. Convergence towards a single fixed point indicates entrainment. The color gradient shows the time at which the Poincare section was taken. B) Comparison of the phase of entrainment fitted to the experimental data between the Dirac and the Gaussian pulse. C) Top: Example of unentrained Gaussian pulse system for a softer pulse (left) and a harder pulse (right). Bottom: Reproduction of figure \ref{fig:fig10}B for a soft and a hard pulse. D-E) Reproduction of the results of figure \ref{fig:fig5} with the Gaussian pulse instead of the Dirac pulse. Demonstration of hysteresis in the Gaussian pulse system with corresponding stroboscopic maps (right). The color gradient indicates the time at which the Poincare section was taken. \label{fig:figS9}}
\end{figure}

Previously, we have used a discontinuous "Dirac" pulse within the pulsatile memory model. This is partly for mathematical convenience. However, such an artificial discontinuity might not be realistic in nature. To show that the results obtained thus far do not rely on such a discontinuity, we reproduce our results using a "softer" pulse. To do so, we use a normalized gaussian pulse acting on the memory variable. This gaussian pulse is "triggered" the same way as the discontinuous version: every completed cycle, at some phase $\phi_0$, the Gaussian pulse is triggered with a constant timescale i.e. the Gaussian pulse goes through a pulse at the same speed regardless of the frequency of the $\phi$ oscillator. Mathematically, we model this as

\begin{equation}\label{eq:soft_pulse}
    \begin{cases}
      \dot{\phi} =& 2\pi f(x) \\
      \dot{\Tilde{t}} = & 1.0\ \textbf{ (if $\phi \mod 2 \pi = \phi_0$, $\ \Tilde{t}=0$)}\\
      \dot{x} =& \alpha \left( r\ g \left(\Tilde{t}; w\right) -x\right), 
    \end{cases} 
\end{equation}

where $g$ is a normalized gaussian centered at $0$,$w$ is the width of the gaussian peak and $\Tilde{t}$ plays the role of the constant timescale. The system's behavior approaches the discontinuous Dirac system as $w \to 0$. This system allows for entrainment, and intrinsic period adaptation as illustrated in figure \ref{fig:figS9}.

\section{Intuition and summary of the behind the models}

\begin{figure*}
\includegraphics[width=18cm]{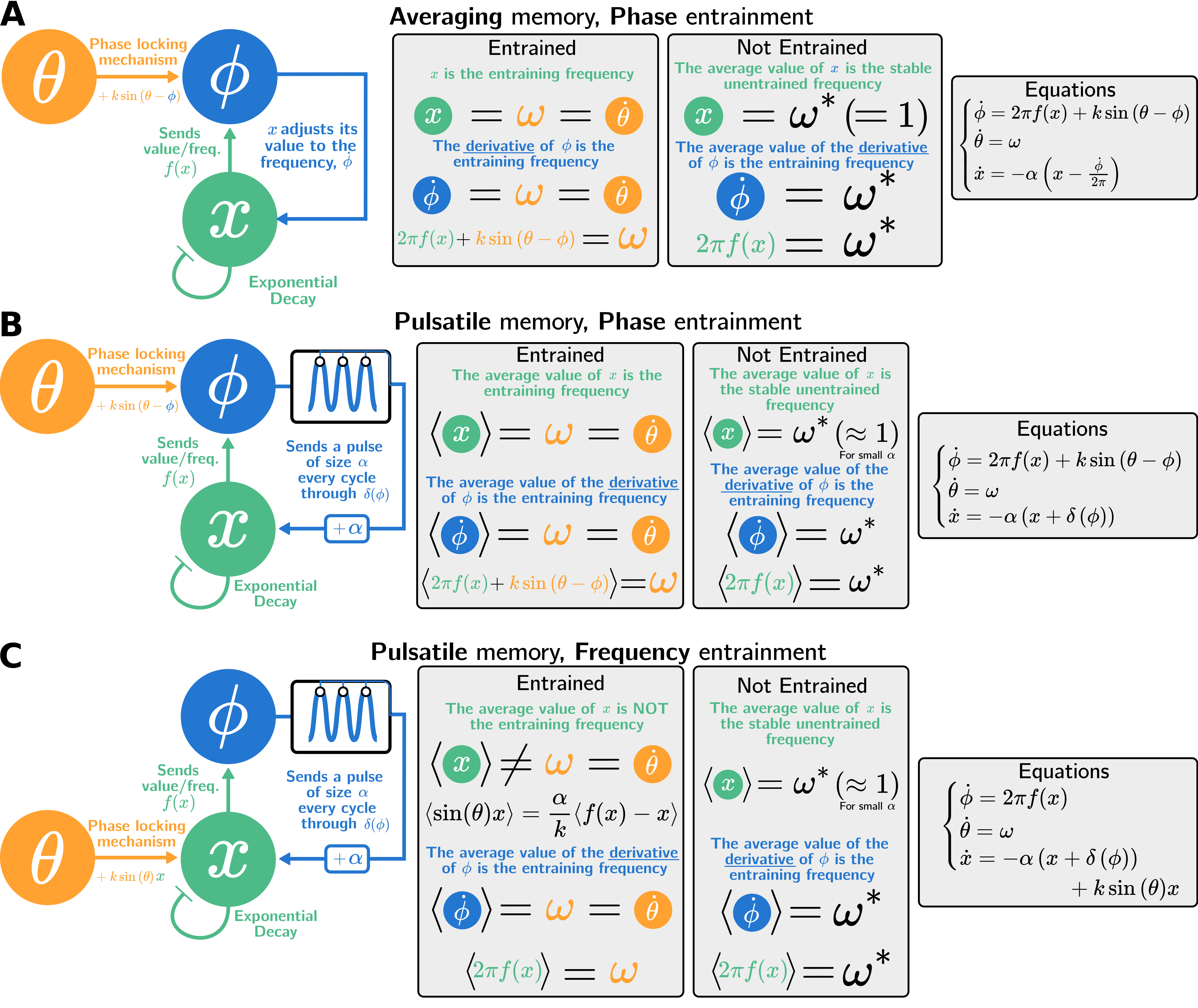}% Here is how to import EPS art
\caption{\label{fig:figS10} A), B), and C) provide intuitive pictures for, respectively, the averaging memory-core phase entrainment system, the pulsatile memory-core phase entrainment system, and the pulsatile memory-frequency entrainment system. A summary of some intuitive insights for both the entrained and not-entrained cases is provided for each system. Notice that B and C have the same "Not Entrained" panel.}
\end{figure*}

Fig. \ref{fig:figS10} summarizes the different models presented in this paper and highlights some key analytical properties about them.

%\bibliography{Biblio}

\end{document}